\documentclass[epj]{svjour}
\usepackage{graphics}
\usepackage{amsmath}
\usepackage{wasysym}
\usepackage[dvipsnames]{xcolor}
\usepackage{physics}

\begin{document}
\title{Beta-decay studies for applied and basic nuclear physics}
\author{
A. Algora \inst{1,2} \and
J. L. Tain \inst{1}  \and
B. Rubio \inst{1} \and
M. Fallot \inst{3} \and
W. Gelletly \inst{4}
}                     
%
\institute{
IFIC (CSIC-Univ. Valencia), Paterna, Spain \and 
Institute of Nuclear Research (ATOMKI), Debrecen, Hungary \and
Subatech (CNRS/in2p3 - Univ. Nantes - IMTA), Nantes, France \and 
University of Surrey, Surrey, UK
}
\date{Received: date / Revised version: date}
%
\abstract{
In this review we will present the results of recent beta-decay studies using the total absorption technique that cover topics of interest for applications, nuclear structure and astrophysics. The decays studied were selected primarily because they have a large impact on the prediction of a) the decay heat in reactors, important for the safety of present and future reactors and b) the reactor electron anti-neutrino spectrum, of interest for particle/nuclear physics and reactor monitoring. For these studies the total absorption technique was chosen, since it is the only method that allows one to obtain beta decay probabilities free from a systematic error called the Pandemonium effect. 
The total absorption technique is based on the detection of the gamma cascades that follow the initial beta decay. For this reason the technique requires the use of calorimeters with very high gamma detection efficiency. The measurements presented and discussed here were performed mainly at the IGISOL facility of the University of Jyv\"askyl\"a (Finland) using isotopically pure beams provided by the JYFLTRAP Penning trap.  Examples are presented to show that the results of our measurements on selected nuclei have had a large impact on predictions of both the decay heat and the anti-neutrino spectrum from reactors. Some of the cases involve beta-delayed neutron emission thus one can study the competition between gamma- and neutron-emission from states above the neutron separation energy. The gamma-to-neutron emission ratios can be used to constrain neutron capture (n,$\gamma$) cross sections for unstable nuclei of interest in astrophysics.
The information obtained from the measurements can also be used to test nuclear model predictions of half-lives and Pn values for decays of interest in astrophysical network calculations. These comparisons also provide insights into aspects of nuclear structure in particular regions of the nuclear chart.
} 
\maketitle
\section{Introduction}
\label{intro}
Our knowledge of the properties of atomic nuclei is derived almost entirely from studies of nuclear reactions and radioactive decays. The ground and excited states of nuclei exhibit many forms of decay but the most common are alpha, beta and gamma-ray emission. Our focus here is on beta decay in its various manifestations. A glance at the Segre Chart reveals that it is the most common way for the ground states of nuclei to decay and it is frequently the observation of such beta decays that brings us our first knowledge of a particular nuclear species and its properties.

The study of beta decay is intrinsically much more difficult than the study of either alpha or gamma decay.  
The reason for this is straightforward. Alpha particles and gamma rays are emitted with discrete energies determined by the differences in energy between the initial and final states involved. Thus characteristic alpha and gamma ray spectra exhibit a series of discrete lines. It requires sophisticated detection and analysis techniques to determine the excitation energies of the states involved, their lifetimes and the transition rates between states. Beta decay carries the same information, but the difficulties of measurement and interpretation are compounded because the spectrum is continuous, not discrete. In 1930 this was explained by Pauli's hypothesis \cite{Pauli} of the existence of a neutral, zero mass particle called in his letter {\it the neutron} that is emitted with the beta particle. The sharing of momentum and energy then explains the continuous spectrum. Shortly afterwards Fermi \cite{Fermi} was able to formulate a theory of beta decay based on this idea and coined the name {\it neutrino } (little neutral one) for the particle. 

\begin{figure}
\resizebox{0.48\textwidth}{!}{ \includegraphics{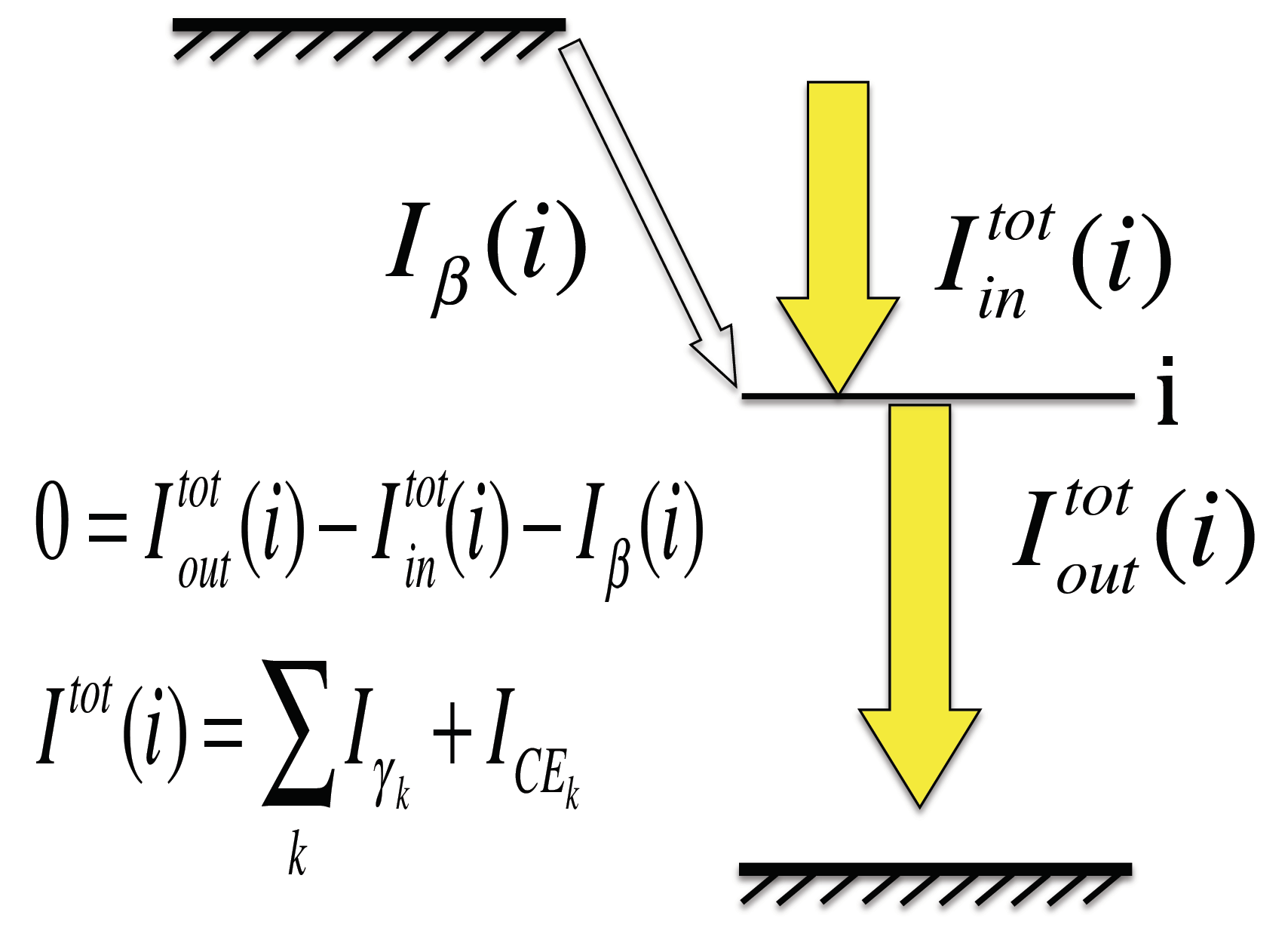}}
\caption{Schematic picture of how the beta feeding is determined in a beta decay experiment employing Ge detectors. The beta feeding (I$_{\beta }(i)$) to level i is determined from the difference of the total intensity feeding the level and those de-exciting it. The sum over (k) represents all transitions feeding or de-exciting the level. I$_{\gamma_{k}}$ stands for the gamma intensity of transition k and I$_{CE_{k}}$ represents the conversion electron intensity. }
\label{balance}       
\end{figure}

A knowledge of beta decay transition probabilities is of particular importance for application to a) tests of nuclear model calculations, b) the radioactive decay heat in reactors, c) the reactor electron anti-neutrino spectrum and d) reaction network calculations for nucleosynthesis in explosive stellar events. In this article we will provide examples of our recent studies of beta decays that involve the use of total absorption gamma spectroscopy (TAGS) to tackle the topics listed above. The TAGS method was adopted in our measurements because it overcomes the difficulties inherent in the conventional use of Ge detector arrays for this purpose. Such arrays are an important and essential tool for constructing nuclear decay schemes since they are very well suited to the study of gamma-gamma coincidences, the main basis for building such schemes. The normal practice is then to derive beta decay transition probabilities for each level populated from the difference in the total intensity of all the gamma rays feeding the level and the sum of the intensities of all those de-exciting it, corrected by the effect of internal conversion (see Figure ~\ref{balance}). In principle this allows us to obtain the beta branching to every level, assuming that we are able to determine by some other means the number of decays that go directly to the daughter ground state, which are not accompanied by gamma emission. 

\begin{figure}
\resizebox{0.48\textwidth}{!}{ \includegraphics{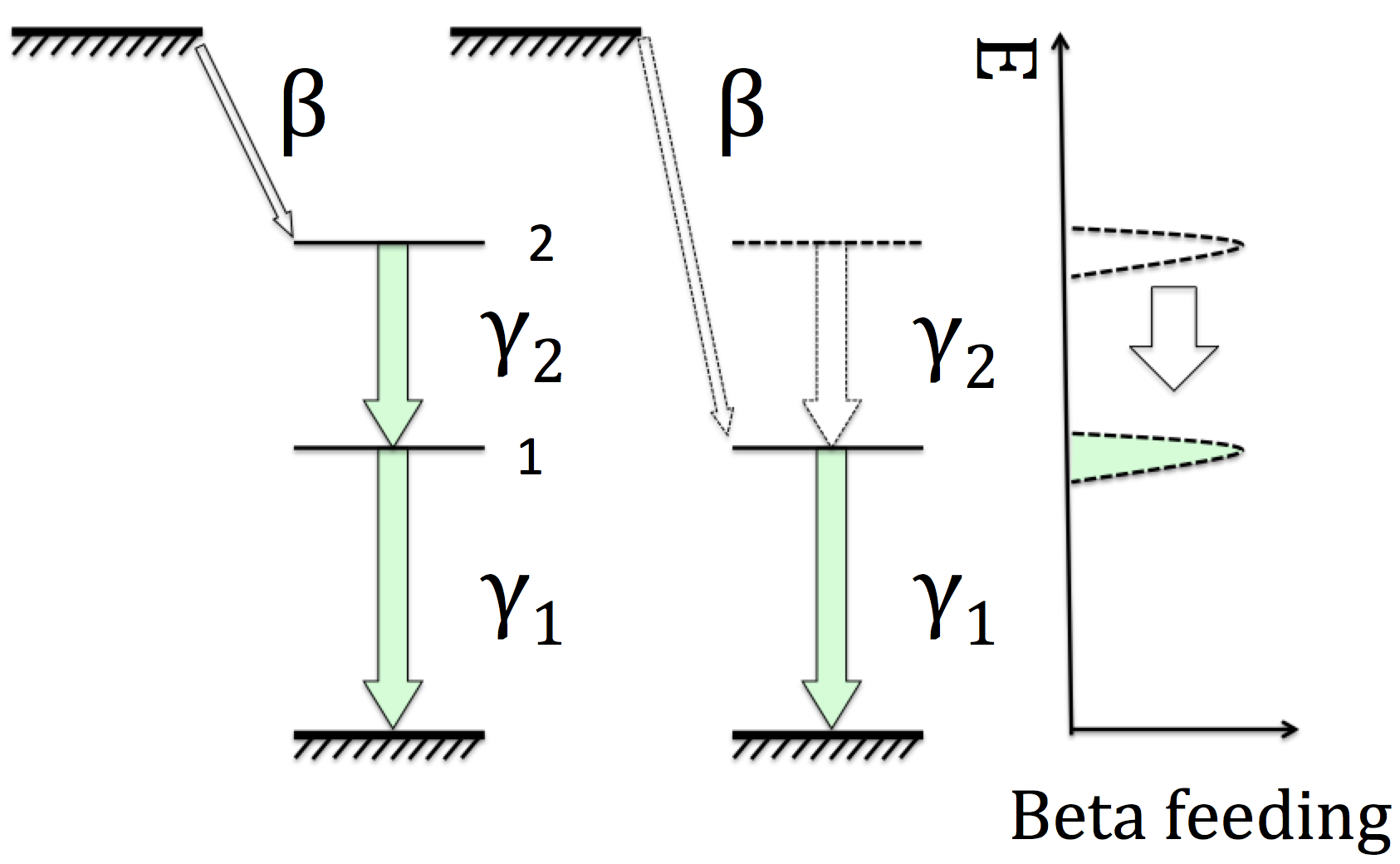}}
\caption{Simplified picture of a beta decay where only one excited state is populated and it de-excites by the emission of a gamma cascade. The left hand panel represents the case. The central panel presents the Pandemonium effect, in this example represented by missing, or not detecting the gamma transition $\gamma_2$. The right hand panel represents the displacement of the beta decay intensity because of the non detection of the transition $\gamma_2$.}
\label{pandemonium}       
\end{figure}

Unfortunately this "simple" procedure does not necessarily give us the correct answers. States at high excitation energies in the daughter nucleus can be populated if the $Q_{\beta}$ value of the decay is large. In this case both the number of levels that can be directly populated by the beta decay is large and the number of levels available to which they can gamma decay is also large. As a result, in general, individual gamma rays      (emitted by levels at high excitation energy) have low intensity. Ge detectors, indeed even gamma-ray arrays, have limited detection efficiencies particularly at higher energies and thus weak transitions are often not detected in experiments. It is clear that this means that we have a problem that has become known as the Pandemonium effect \cite{Hardy} (see Figure ~\ref{pandemonium} for a simplified picture). 

We can overcome this problem using the total absorption gamma spectroscopy technique, where we take a different approach. The method involves a large 4$\pi$ scintillation detector and is based on the detection of the full de-excitation gamma cascade for each populated level, rather than the individual gamma rays. The power of TAGS to find the missing beta intensity has been demonstrated in a number of papers \cite{Hu_98,Nacher,Poirier,Algora,Jordan,Perez,Briz,Aguado,Tain,Guadilla2019}. The use of the TAGS method began at ISOLDE\cite{Duke}. Its development and history are described in \cite{Rubio2005,Rubio}.


Looking at a wider picture we see that many entries in the international databases, that rely on measurements with Ge detectors alone, will have systematic errors. As we shall see in the sections that follow this means that the results cannot be relied on for certain applications. The answer to the resulting difficulties lies in the use of TAGS. In the remainder of this article we will describe the TAGS method in more detail and then use our results to illustrate how it can be applied.

The structure of this article is the following: in Section ~\ref{section_TAS} details of the experimental method and the analysis of the spectra 
are described. Sections ~\ref{decay_heat}, ~\ref{neutrino}, ~\ref{nuclear} and ~\ref{astrophysics} deal with beta decay studies related to  a) radioactive decay heat (DH), b) reactor antineutrino spectra c) nuclear models, and d) astrophysical applications respectively.
Finally, in Section ~\ref{summary_future}, a summary  will be presented.

\section{TAGS measurements}
\label{section_TAS}
In Section~\ref{intro} it was already explained why we need TAGS measurements. Figure ~\ref{different_detectors} shows how the simple beta decay presented in Figure ~\ref{pandemonium} is detected by typical detectors used in beta decay experiments. Because a TAGS detector acts like a calorimeter, in an ideal TAGS experiment the detected spectrum will be proportional to the beta intensity distribution. This spectrum is obtained in ideal conditions, where there is no penetration of the beta particles, or the radiation generated by them, into the detector and the TAGS detector is 100$\%$ efficient up to the full energy of the gamma rays that follow the beta decay. That means that only the full absorption peak corresponding to the sum energy of the gamma cascade is detected in the case of a $\beta^-$ decay.  

A real experiment does not quite match this ideal. In order to achieve very high detection efficiencies, large, close to $4\pi$, detector volumes are needed. Thus inorganic scintillation material has been the natural choice. Because of its good average properties NaI(Tl) has been used in all except one (see later) of the existing spectrometers. Nevertheless  we need some opening to take the sources to the centre of the spectrometer, either in the form of a radioactive beam or deposited
onto a tape transport system. The latter may also be needed to remove the sources after some measuring time. We may also need ancillary detectors for detecting coincidences and selecting the events in which we are interested. In addition TAGS detectors require, in general, some form of encapsulation. All these requirements mean that we have dead material and holes in our detector system. Accordingly the gamma detection efficiency of our system will not be 100$\%$. The consequence is that to obtain the beta intensity distribution we need to solve the inverse problem represented by the following equation:  

\begin{figure}
\resizebox{0.50\textwidth}{!}{ \includegraphics{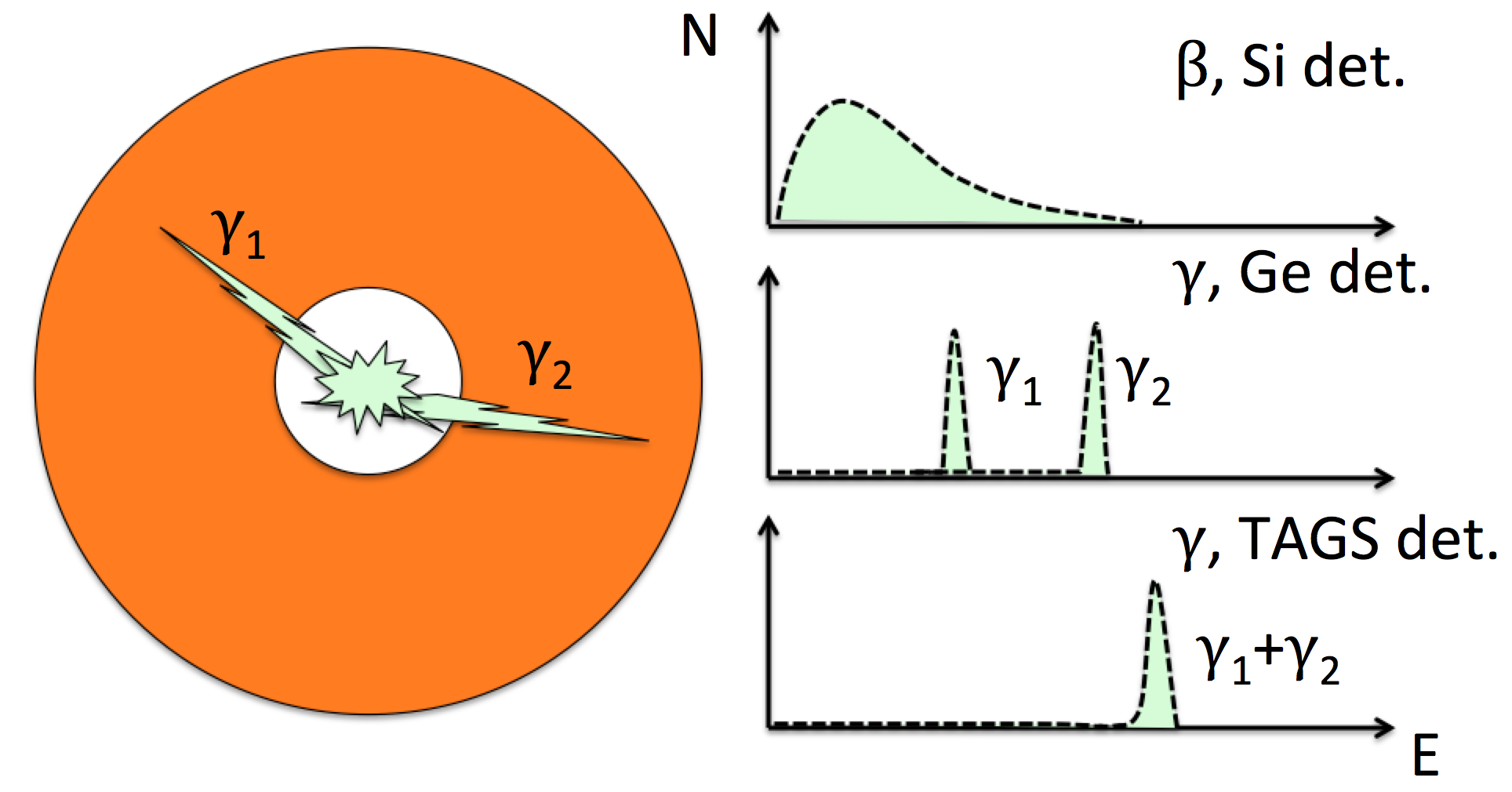}}
\caption{Schematic picture of how the simple beta decay depicted in Figure ~\ref{pandemonium} is seen ideally by different detectors used in beta decay studies. Left panel, representation of a total absorption detector, rigth panel, ideally detected spectra with a beta dectector (a silicon detector), a Ge detector and a total absorption detector after the simple decay represented in Figure ~\ref{pandemonium}.}
\label{different_detectors}       
\end{figure}

\begin{equation}\label{response}
  d_i = \displaystyle\sum_{j=0}^{j_{max}}R_{ij}(B)f_j + C_i
\end{equation}

where $d_i$ is the content of bin $i$ in the  measured TAGS spectrum, $R_{ij}$ is the response matrix of the TAGS setup and represents the probability that a decay that feeds level $j$ in the level scheme of the daughter nucleus gives a count in bin $i$ of the TAGS spectrum, $f_j$ is the beta feeding to the level $j$ (our goal) and $C_i$ is the contribution of the contaminants to bin $i$ of the TAGS spectrum. The index $j$ in the sum runs  over the levels populated in the daugther nucleus in the beta decay. The response matrix $R_{ij}$ depends on the TAGS setup and on the assumed level scheme of the daughter nucleus. The dependence on the level scheme of the daughter nucleus is introduced through the branching ratio matrix $B$. This matrix contains the information of how the different levels in the assumed level scheme decay to the lower lying levels. To calculate the response matrix $R_{ij}(B)$ the branching ratio matrix $B$ has to be determined first. 
There are different ways to extract the feeding distribution from equation ~\ref{response} or, in other words,  to solve the TAGS inverse problem. One can assume the existence of "pseudo" levels that are added manually (with their decaying branches) to the known level scheme, calculate their response and see their effect in the calculated spectrum (see for example \cite{Greenwood,Polish}). In our analysis until now we have followed an alternative way for which the level scheme of the daughter nucleus is divided into two regions, a low excitation part and a high excitation part. Conventionally the levels of the low excitation part and their gamma decay branchings are taken from high resolution measurements available in the literature, since it is assumed that the gamma branching ratios of these levels are well determined. Above a certain energy, the cut-off  energy, a continuum of possible levels divided into 40 keV bins is assumed.  From this energy up to the  decay $Q_\beta$ value, the statistical model is used to generate a branching ratio matrix for the high excitation part of the level scheme. The statistical model is based on a level density function and gamma strength functions of E1, M1, and E2 character. 
Once the branching ratio matrix (B) is defined, the response of the setup $R_{ij}$ to that branching matrix B (or level scheme) is calculated using previously validated Monte Carlo simulations of the relevant electromagnetic interactions in the experimental setup. The validation of the Monte Carlo simulations is performed by reproducing  measurements of well known radioactive sources, that are made under the same experimental conditions as the real experiment. The Monte Carlo simulations require a careful implementation of all the details of the geometry of the setup, a proper knowledge of the materials employed in the construction of the setup and testing to find the best Monte Carlo tracking options and physics models that reproduce the measured sources. It should be noted that from high resolution measurements we use only the branching ratios of the levels, and not the information on the feeding of these levels. 

Once the response function is determined we can solve Equation ~\ref{response} using appropriate algorithms to determine the feeding (or beta intensity) distribution. In our analyses we follow the procedure developed by the Valencia group. In \cite{Tain_analysis} several algorithms were explored. From those that are possible, the expectation maximization (EM) algorithm is conventionally used, since it provides only positive solutions for the feeding distributions and no additional regularization parameters (or assumptions) are required to solve the TAGS inverse problem.

Clearly, the first  level scheme (or defined branching ratio matrix) considered is not necessarily the one that will provide a nice description of the measured TAGS spectrum. For that reason, as part of the analysis the cut-off energy and the parameters that define the branching ratio matrix can be varied until the best description of the experimental data is obtained. Also assumptions on the spin and parity of the ground state of the parent nucleus and on the spins and parities of the populated levels can be changed when they are not known unambigously, since to connect the levels in the continuum to levels in the known part of the level scheme we need information about their spin and parity. All these changes provide different branching ratio matrices 
(or daughter level schemes) that are considered during the analysis and for all of them the corresponding response matrixes are calculated and  Equation ~\ref{response} is solved. The final analysis is then based on the level scheme (or branching ratio matrix) that is consistent with the available information from high resolution measurements and at the same time provides the best description of the experimental data. So in practical terms the following steps are followed until the best description of the data is obtained: a) define a branching ratio matrix $B$, b) calculate the corresponding response matrix 
$R_{ij}(B)$, and c) solve the corresponding Equation ~\ref{response} using an appropriate algorithm d) compare the generated spectrum after the analysis ($R(B)f+C$) with the experimental spectrum $d$. 

We have only mentioned briefly how the response function $R_{ij}(B)$ is calculated. More specifically the response for each level can be determined recursively starting from the lowest level in the following way \cite{Cano_response}:

\begin{equation}\label{response2}
R_j= \displaystyle\sum_{k=0}^{j-1}b_{jk}g_{jk} \otimes R_k
\end{equation}

where $R_j$ is the response to level $j$, $g_{jk} $ is the response of the gamma transition from level $j$ to level $k$ which is calculated using Monte Carlo simulations, $b_{jk}$ is the branching ratio for the gamma transition connecting level $j $ to level $k$, and $R_k$ is the response to level $k$. Here the index $k$ runs for all the levels below the level $j$. For simplicity we have not included here in the formula the convolution with the response of the beta particles and only the gamma part of the response is presented. Note that in this last notation $R_j$ is a vector that contains as elements the $R_{ij}$ matrix elements mentioned above for all possible $i$-s (or channels) of the TAGS spectrum and the branching ratio matrix enters in the formula of the response matrix through the decay branches $b_{jk}$-s. In the real calculation of the responses the internal conversion process is also taken into account. 

Prior to the analysis, the contaminants in the TAGS spectrum ($C_i$) have to be isolated and their individual contributions evaluated.
The nucleus to be studied is produced by nuclear reactions together with a number of additional nuclei. 
Two alternative separation methods are normally used to isolate the nucleus of interest. On-line mass separators are used with low energy radioactive beams to reduce the contamination represented by mass isobars. In-flight separation is used at high-energy fragmentation facilities to reduce the number of  nuclear species in the ''cocktail beam" to suitable levels. Even if we can isolate the nucleus of interest,  daughter (and other descendants) activity can contaminate the measured spectrum depending on the half-life of the studied decay. This contamination can be determined through dedicated measurements on the decay of the contaminant nuclei under the same conditions as the one of interest. Another source of contamination of the spectrum is the pile-up of signals. The pile-up can distort the full TAGS spectrum and  can generate counts in regions of the spectra where there should not be counts, as for example in the region beyond the $Q_\beta$ value of the decay. Also it can distort the spectrum in regions where we expect reduced statistics as for example close 
to the $Q_\beta$ value of the decay. This is the reason why estimating this contribution is of importance. Algorithms have been developed to evaluate this contribution \cite{pileup,Guadilla_nim}.
Its determination is based on the random superposition of true detector pulses, measured during the experiment, within the time interval defined by the acquisition gate of the data acquisition system. 

\begin{figure}
\resizebox{0.48\textwidth}{!}{ \includegraphics{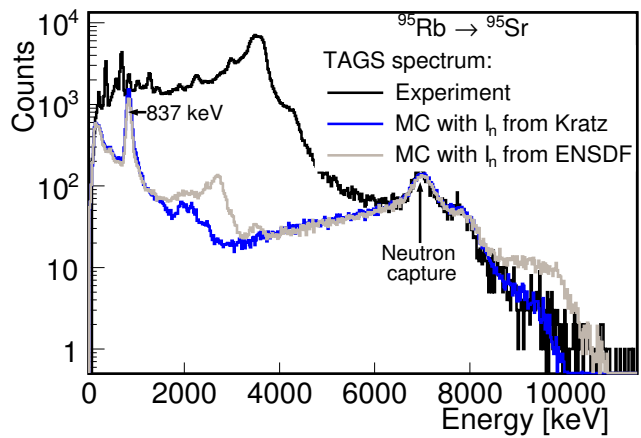}}
\caption{Impact of the neutron energy spectrum ($I_n$) in the simulations of the contamination associated with the beta delayed neutrons in the TAGS spectrum (for more details see \cite{Guadilla_neutron}). Only the spectrum measured by Kratz {\it et al.} \cite{Kratz1} reproduces the TAGS spectrum at high excitation energies. The Monte Carlo (MC) spectra are normalized to the experimental spectrum around the neutron capture peak indicated with an arrow. The prompt 836.9 keV $\gamma$-ray peak from the first excited state in the final nucleus after the beta-delayed neutron emission $^{94}$Sr is highlighted. Reprinted figure with permission from \cite{Guadilla_neutron}, Copyright (2019) by the American Physical Society.}
\label{neutron_contamination}       
\end{figure}

Another possible contamination appears when the decay is accompanied by beta delayed particle emission, since this process can lead promptly to the emission of gamma rays from the final nucleus populated by the  beta delayed particle emission.  The case of the emission of beta delayed neutrons is even more complex. Neutrons interact easily with the detector material and release their energy through inelastic and capture processes. The proper evaluation of this contamination is of great relevance in the study of beta decays far from stability on the neutron-rich side of the Segre chart and requires careful Monte Carlo simulations of the neutron-detector interactions \cite{Guadilla_nim,Tain_nsens}. The reproduction of this contamination is complicated because it has two components: one,  which is prompt with the beta decay, is composed of gamma rays emitted in the final nucleus after the beta-delayed neutron emission when an excited state is populated, the other component due to neutron interactions in the detector is delayed, since the speed of neutrons is much lower than that of gamma-rays. To simulate these effects properly an event generator \cite{Valencia}, that takes into account relative contribution of the two components is required. It is also necessary to know the energy spectrum of the emitted beta-delayed neutrons. In addition,  
the Monte Carlo simulation code should include an adequate physics model of the neutron interactions.
As an example, in Figure \ref{neutron_contamination} \cite{Guadilla_neutron} the contribution of the calculated beta delayed neutron contamination to the TAGS decay spectrum of $^{95}$Rb is presented. Two available neutron energy spectra were used in the simulations \cite{Kratz1,Abriola1}, and clearly only one reproduces the experimental TAGS data at high excitation energies. This figure shows the relevance of the neutron spectrum used in the simulations (for more details see \cite{Guadilla_neutron}). 
Due to these complications we have built {\it Rocinante} \cite{Valencia,Tain_nd2010} a spectrometer made of BaF$_{2}$ material, aimed at the measurement of beta-delayed neutron emitters. BaF$_{2}$ has a neutron capture cross-section one order-of-magnitude smaller than the NaI(Tl), that is conventionally used. This spectrometer was also the first of a new generation of segmented devices designed to exploit the cascade multiplicity information to improve the TAGS analysis, as will be mentioned later.

It is important to first identify the different distortions or contaminations, but it is also important to determine properly their corresponding weight in the measured spectrum. Depending on the distortion, different strategies have been followed. For example, the contribution from contaminant decays can be evaluated if there is a clear peak identified in the spectrum that comes from this contamination that can be used for normalization. Another option is the assessment of this contribution from the solution of the Bateman equations, using the information on half-lives and measurement conditions (collection and measuring cycle times). In the case of high-energy fragmentation experiments where the contamination is due to beta-gamma events uncorrelated with the implanted ion it can be evaluated from correlations backward in time. The pileup distortion can be evaluated based on the number of counts in the TAGS spectra which lie beyond the highest $Q_\beta$ value in the decay chain and which are clearly above the contribution of the background, since we can assume that those counts can only come from this contribution. When this option is not possible because of inadequate statistics,  a procedure is given in \cite{pileup} for the normalisation of this contribution based on the counting rate and the length of the 
analogue to digital converter (ADC) gate. And finally if there is a contamination arising from beta-delayed neutrons, this contribution can be normalized to the broad high-energy structure generated by neutron captures in the detector material when possible, otherwise it should be normalized to the Pn value of the decay. 

\begin{figure}
\resizebox{0.50\textwidth}{!}{ \includegraphics{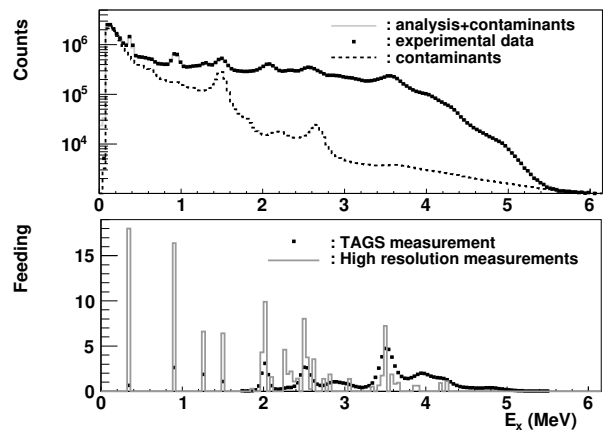}}
\caption{Comparison of the measured TAGS spectrum of the decay of $^{104}$Tc with the spectrum generated after the analysis (reconstructed spectrum). This last spectrum is obtained by multiplying the response function of the decay with the determined feeding distribution ($R(B)f^{final}$). The lower panel shows the beta-decay feeding distribution obtained compared with that previously known from high resolution measurements \cite{Algora,Jordan,Jordan-thesis}. Reprinted figure with permission from \cite{Algora}, Copyright (2010) by the American Physical Society.}
\label{104Tc}       
\end{figure}

In Figure \ref{104Tc}, we present as an example a total absorption spectrum measured during our first experiment in Jyv\"askyl\"a of the decay of $^{104}$Tc \cite{Algora,Jordan} which is relevant for the decay heat application (see Section \ref{decay_heat}). In the upper panel of this figure we show the spectrum of this decay compared with the reproduction of the spectrum after the analysis and the contribution of the contaminants (background+daughter activity+pileup). 

In this measurement a  TAGS detector that consisted of two NaI(Tl) cylindrical crystals with dimensions: $\diameter = 200\,\mathrm{mm} \times l = 200\, \mathrm{mm}$, and $\diameter = 200\, \mathrm{mm} \times l = 100\, \mathrm{mm}$ was used (courtesy of Dr. L. Batist). The longer crystal has a longitudinal hole of $\diameter = 43\, \mathrm{mm}$ for the positioning of the sources in the approximate geometrical centre of the spectrometer using a tape system. In the experiment the crystals were separated by 5~mm. This separation and the ideal position of the sources inside the spectrometer was studied previous to our experiment using Monte Carlo simulations in order to maximize the gamma efficiency of the setup \cite{Algora-annualreport}. This TAGS had a 57\% peak and 92\% total efficiency for the 662 keV gamma transition emitted in $^{137}$Cs decay and 27\% peak and 70\% total efficiency for a 5 MeV gamma transition. This last value was obtained from previously validated Monte Carlo simulations. The efficiency of this setup is modest compared with recently developed total absorption spectrometers such as DTAS \cite{DTAS}, or MTAS \cite{MTAS}. This detector was designed at the Nuclear Institute of St. Petersburg (Russia) \cite{Leonid}. 

This measurement was analyzed in singles, since the precision and reproducibility of the tape positioning system was considered not sufficiently good to allow coincidence counting. The positioning of the sources is critical in the determination of the efficiency of the Si detector used as an ancilliary detector for coincidences with the beta particles emitted in the decay. The efficiency of the beta detector as a function of end-point energy has a direct impact on the normalization of the combined beta-gamma cascade response of the spectrometer ($R_{ij}$). Using singles has the advantage of providing much higher statistics in the analysis compared with gated spectra. However, the use of gated spectra is preferred, eventually unavoidable, in order to reduce contamination from ambient background and the selection of events of particular interest. 

The lower panel of Figure \ref{104Tc} shows the feeding distribution deduced for the $^{104}$Tc decay obtained from the TAGS measurement compared with the distribution obtained from high resolution measurements. From the figure it is clear that the feeding distribution obtained with the TAGS is shifted to higher energies in the daughter nucleus, which is typical of a case suffering from the Pandemonium effect \cite{Algora,Jordan,Jordan-thesis}. Similar measurements will be discussed in more detail in Section  \ref{decay_heat}. 


As mentioned earlier, in this measurement a detector was used that was composed of two crystals. The new generation of available detectors such as {\it Rocinante} \cite{Valencia}, SUN \cite{SUN}, MTAS \cite{MTAS} and DTAS \cite{DTAS} exploit segmentation to a greater degree to extract additional information from 
the decay under study. Using the segmentation it is possible to measure the detector fold (number of detectors fired in an event) which
is related to gamma-cascade multiplicity as a function of excitation energy and ultimately to the de-excitation branching ratio matrix $B$. The lack of knowledge of the matrix $B$ is the largest source of uncertainty in TAGS analysis and this can be greatly improved with segmented detectors. Our current  approach to the iterative procedure for updating $B$ described earlier in this Section, is to include in step d) the comparison to fold-gated TAGS spectra and single module spectra. Reconstructed fold-gated spectra are obtained by MC simulation using the appropriate event generator since it is not possible to define a fold-gated response in a manner similar to Equation \ref{response2}. This also prevents us from including them as part of the inverse problem (Equation \ref{response}).  
A different approach has been taken by the ORNL group to analyze MTAS data \cite{Rasco,Rasco_JPS}. They use the coincidence between one module and the sum of all the modules to define total energy gated single detector spectra that are fitted by the sum of a number of de-excitation cascades, usually taken from high resolution spectroscopy and supplemented when necessary with "pseudo" levels with guess branching ratios and modified iteratively until the best reproduction is achieved.  
Yet another approach is used by the NSCL group to extract $B$ from SUN data \cite{Spyrou_Oslo}. They start from the same total energy gated single detector spectra but apply the so called Oslo-method \cite{Oslo} to obtain the branching ratio matrix for a subset of levels. Because of this, the TAGS analysis is not performed with this $B$ but uses the "pseudo" level approach including in the fit the total absorption spectrum and the spectrum of detector multiplicities \cite{Dombos}. It should be noticed that the traditional Oslo method is not strictly applicable to TAGS data because the assumed equivalence of total deposited energy with excitation energy does not hold in general, due to the non-ideal detector response.  Currently we are working in a method to solve the full non-linear inverse problem represented by Equation~\ref{response}, to obtain feedings and branching ratios from the complete data set provided by a segmented spectrometer: sum energy spectrum gated by detector fold, sum energy spectrum versus single crystal spectrum and crystal-crystal correlations.

\begin{figure}
\resizebox{0.50\textwidth}{!}{ \includegraphics{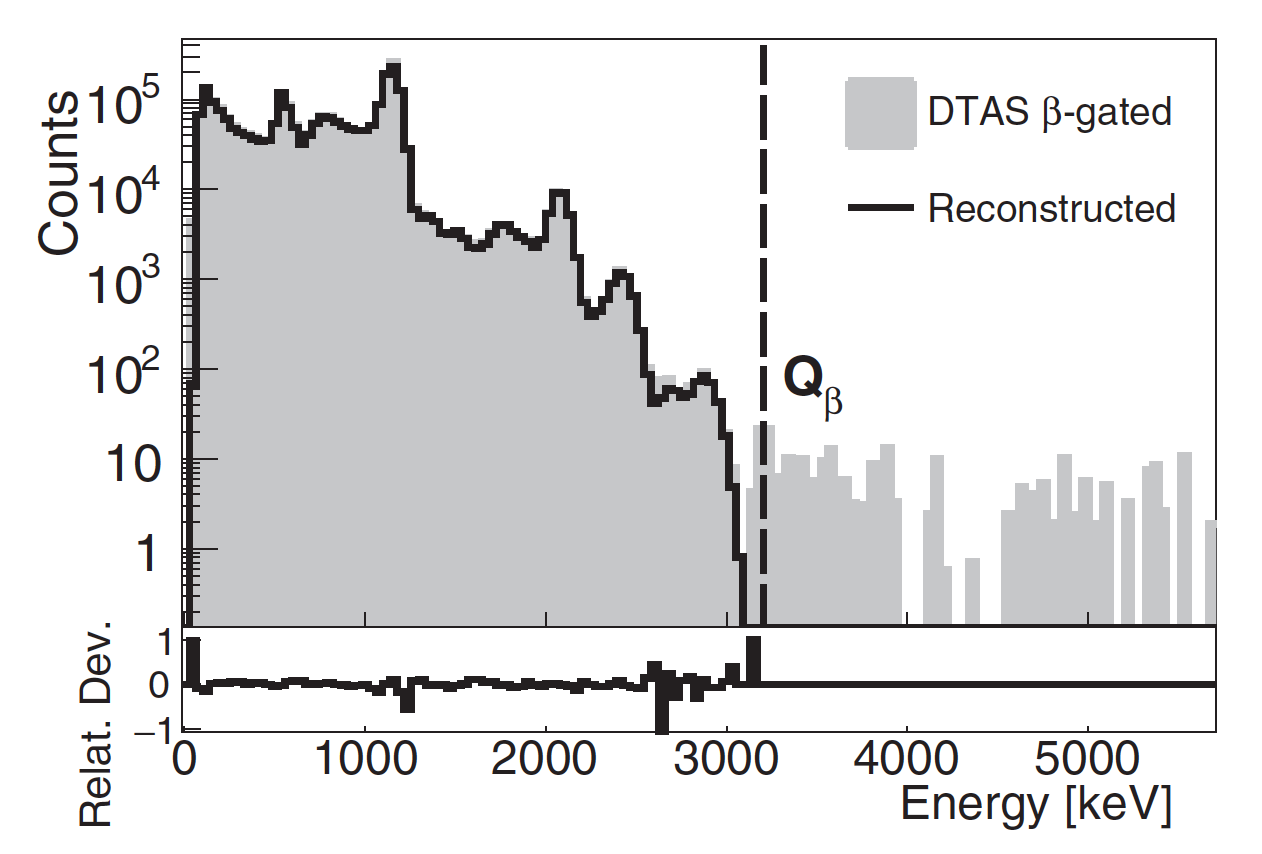}}
\caption{Comparison of the measured TAGS spectrum of the decay of $^{100}$Tc with the spectrum generated after the analysis (reconstructed spectrum).
Reprinted figure with permission from \cite{100Tc}, Copyright (2017) by the American Physical Society. }
\label{100Tc}       
\end{figure}
\begin{figure}
\resizebox{0.50\textwidth}{!}{ \includegraphics{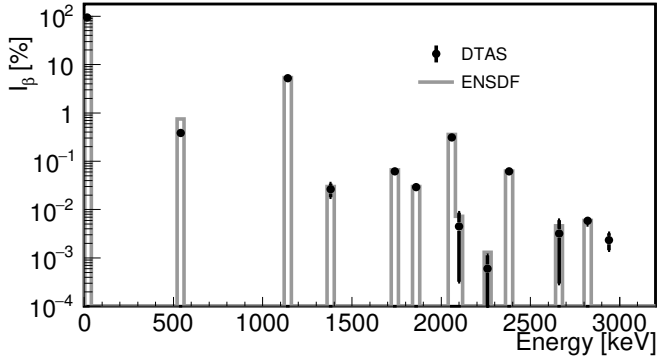}}
\caption{Comparison of the obtained TAGS feeding distribution in the decay of $^{100}$Tc with the data available from high resolution measurement. This is an example of a case that did not suffer from the Pandemonium effect. Reprinted figure with permission from \cite{100Tc}, Copyright (2017) by the American Physical Society.}
\label{100Tc_feeding}       
\end{figure}

In Figure \ref{100Tc} we show the spectrum of the beta decay of $^{100}$Tc measured in a recent campaign of measurements at the IGISOL IV facility of the Univ. of Jyv\"askyl\"a \cite{Guadilla-thesis,100Tc}  with the segmented DTAS detector. This single decay is part of the A=100 system of relevance for double beta decay studies ($^{100}$Ru - $^{100}$Tc - $^{100}$Mo). Previous to this study, only a high resolution measurement existed for this single decay and there were doubts whether feeding at high excitation energy is not detected in the high resolution measurements. Single decays, like this one can be of relevance for fixing model parameters used in theoretical calculations for neutrino and neutrinoless double beta decay studies. Our TAGS results show only a 
modest improvement with relation to the earlier high resolution results, revealing that this decay did not suffer seriously from the Pandemonium systematic error (see Figure \ref{100Tc_feeding}). This decay is not only important in the framework of double beta decay studies, it has also recently attracted attention in another neutrino related topic \cite{Huber2}. The decay is a relevant contributor in a
newly identified flux-dependent correction to the antineutrino spectrum produced in nuclear reactors that takes into account the contribution of the decay of nuclides that are produced by neutron capture of long lived fission products. In this particular case $^{99}$Tc is produced as a fission product, which after neutron capture becomes $^{100}$Tc that beta decays. The effect has a nonlinear dependence on the neutron flux, because first a fission is required and later a neutron capture. Effects like this one are considered in order to explain features of the predicted antineutrino spectrum for reactors not yet fully understood (see Section \ref{neutrino}). 
 
\begin{figure}
\resizebox{0.50\textwidth}{!}{ \includegraphics{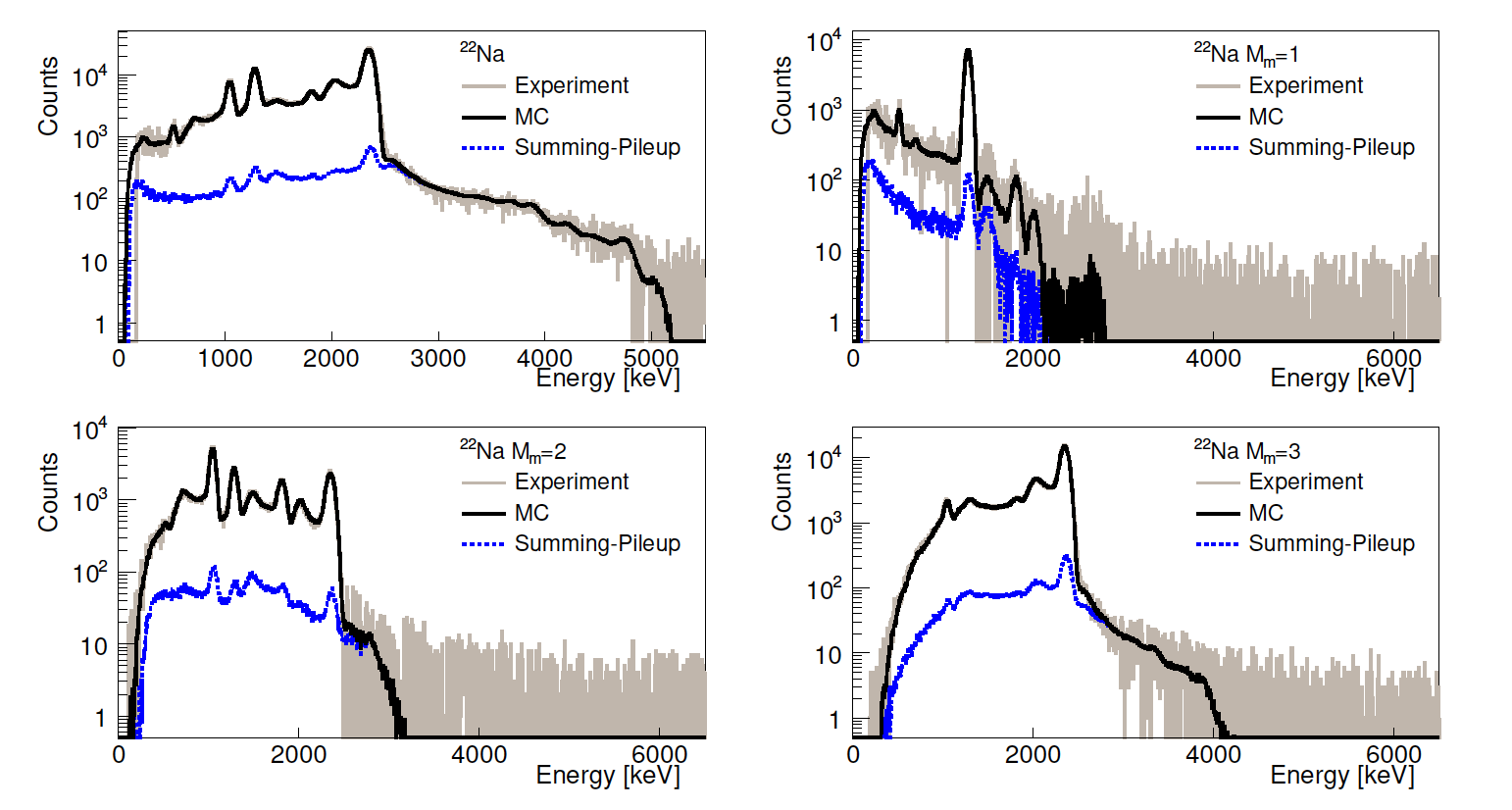}}
\caption{Comparison of the measured spectrum of the decay of $^{22}$Na with different multiplicity conditions on the number of detectors that fired (fold) with the results of Monte Carlo simulations 
for the DTAS detector \cite{Guadilla_nim,Guadilla-thesis}. How well the different multiplicity spectrum is reproduced, is a stringent test of the quality of the branching ratio matrix used in the analysis. Modified figure with permission from \cite{Guadilla_nim}, Copyright (2018) by Elsevier.}
\label{22Na}       
\end{figure}

The study of this decay was the first time that the DTAS detector was used at a radioactive beam facility. Prior  to the analysis of this case a full characterization of the detector was performed \cite{Guadilla_nim,Guadilla-thesis}. This included a check on the ability to reproduce with MC simulations the spectrum of decays obtained with different detector multiplicity (fold) conditions. 
As an example we present in Figure \ref{22Na} the reproduction of the multiplicities for the $^{22}$Na  source used in the characterization of the detector. 
The DTAS is constructed in a modular way that adds extra versatility to the setup \cite{DTAS}. Depending on the installation, it can be used in an 18 detector configuration for ISOL type facilities or in an  16 detector configuration for fragmentation facilities, where the positioning of the implantation detectors normally requires more space. 


\section{Decay heat}
\label{decay_heat}

Nuclear reactor applications require beta decay data. 
The relevance of beta decay is shown by the fact that each fission is followed by approximately six beta decays. The energy balance released in fission is presented in Table \ref{table235U} for $^{235}$U and $^{239}$Pu fissile isotopes \cite{235Uenergy}. In the case of $^{235}$U, for example,  7.4 $\%$ of the energy  released comes from the beta decay of the fission products (FP) (gamma and beta energy). Depending on the composition of the fuel in the reactor this percentage can change, but it is of the order of 7$\%$ of the total released energy for a working reactor. Once the reactor is shut-down, the decay energy becomes dominant and the related heat has to be removed. If for some reason this is not possible, it can produce  accidents like the one caused originally by the tsunami that followed the Great East Japan Earthquake (2011) in the Fukushima Daiichi power plant. Clearly one needs to estimate this source of energy for the safety of any nuclear installation (design of a reactor, storage of the nuclear waste, loss of coolant accident (LOCA), etc.). 

Decay heat is defined as the amount of energy released by the decay of fission products not taking into account the energy taken by the neutrinos. The first method to estimate the decay heat was introduced by Way and Wigner \cite{WayWigner}, which was based on statistical considerations of the fission process. Their results provide a good estimate of the heat released, but the precision reached is not sufficient for present-day safety standards. 
Nowadays the most extended way to estimate the decay heat is to perform summation calculations, which relies on the increased amount of available nuclear data. In this method, the power function of the decay heat $f(t)$ is obtained as the sum of the activities of the fission products times the energy released per decay: 

\begin{equation}
f(t)= \sum_{i} (\overline{E}_{\beta,i} + \overline{E}_{\gamma,i}) \lambda_i N_i(t) 
\label{eq:power-function}
\end{equation}

where $\overline{E}_{i}$ is the mean decay energy of the $ith$ nuclide ($\beta$ or charged-particle and $\gamma$ or electromagnetic  components), $\lambda_i$ is the decay constant of the $i$th nuclide, and $N_i(t)$ is the number of nuclides of type $i$ at the cooling time $t$ (for simplicity the $\alpha$-decays of minor actinides are not included here). These calculations require extensive libraries of cross sections, fission yields and decay data, since the method first requires the solution of a system of coupled differential equations to determine the inventory of nuclei $N_i(t)$ produced in the working reactor and after shut-down. 

\begin{table}
\caption{Division of the energy released by the most important fissile isotopes $^{235}$U and $^{239}$Pu (values given in MeV/fission) \cite{235Uenergy}.}
\label{table235U}       
\begin{tabular}{lrr}
\hline\noalign{\smallskip}
Contribution & $^{235}$U & $^{239}$Pu \\
\noalign{\smallskip}\hline\noalign{\smallskip}
Fragments' kinetic energy & 166.2(13) & 172.8(19)\\
Prompt neutrons & 4.8(1) & 5.9(1)\\
Prompt gamma rays & 8.0(8) & 7.7(14)\\
Beta energy of fission fragments & 7.0(4)&6.1(6)\\
Gamma energy of fission fragments  & 7.2(13) & 6.1(13) \\
\hline\noalign{\smallskip}
Subtotal & 192.9(5) & 198.5(8)\\
Energy taken by the neutrinos & 9.6(5) & 8.6(7)\\
\hline\noalign{\smallskip}
Total  & 202.7(1) & 207.2(3)\\
\noalign{\smallskip}\hline
\end{tabular}
\end{table}

\begin{table}
\caption{List of parent nuclides identified by the WPEC-25 (Nuclear Energy Agency working group) that should be measured using the total absorption technique to improve the 
predictions of the decay heat in reactors \cite{WPEC25,Nichols1}. These nuclides are of relevance for conventional reactors based on $^{235}$U and $^{239}$Pu fission. The list contains 37 nuclides. Rel. (relevance) stands for the priority of the measurement. Isotopes marked with asterisks show the  measurements performed by our collaboration. 
Nuclides marked with $\dagger$ are also relevant for the $^{233}$U/$^{232}$Th fuel, see additional cases in Table \ref{tab:3}. The isotopes are identified according to the 
Z-Symbol-A notation; m stands for metastable or isomeric state.}
\label{tab:2}       
\begin{tabular}{l c l c l c}
\hline\noalign{\smallskip}
Isotope & Rel. & Isotope & Rel. & Isotope & Rel.\\
\noalign{\smallskip}\hline\noalign{\smallskip}
35-Br-86$^\dagger$$^*$  	& 1  & 41-Nb-99$^\dagger$          & 1 & 52-Te-135$^\dagger$     & 2\\
35-Br-87$^\dagger$$^*$      & 1 & 41-Nb-100$^\dagger$$^*$  & 1 & 53-I-136$^\dagger$        & 1\\
35-Br-88$^\dagger$$^*$      & 1 & 41-Nb-101$^\dagger$$^*$  & 1  & 53-I-136m$^\dagger$    & 1\\
36-Kr-89$^\dagger$             & 1 & 41-Nb-102$^\dagger$$^*$  & 2 & 53-I-137$^\dagger$$^*$ & 1\\
36-Kr-90$^\dagger$             & 1 & 42-Mo-103$^\dagger$$^*$  & 1 & 54-Xe-137$^\dagger$     & 1\\
37-Rb-90m                            & 2 & 42-Mo-105$^*$                    & 1 & 54-Xe-139$^\dagger$     & 1\\
37-Rb-92$^\dagger$$^*$     & 2 & 43-Tc-102$^\dagger$$^*$  & 1 & 54-Xe-140$^\dagger$      & 1\\
38-Sr-89                                & 2 & 43-Tc-103$^\dagger$$^*$  &1  & 55-Cs-142$^*$                 & 3\\
38-Sr-97                                & 2 & 43-Tc-104$^\dagger$$^*$  & 1 & 56-Ba-145                        & 2\\
39-Y-96$^\dagger$              & 2 & 43-Tc-105$^*$                     & 1 & 57-La-143                        & 2\\
40-Zr-99$^\dagger$             & 3 & 43-Tc-106$^*$                     & 1 & 57-La-145                        & 2\\
40-Zr-100$^\dagger$           & 2 & 43-Tc-107$^*$                     & 2\\
41-Nb-98$^\dagger$$^*$    & 1 &  51-Sb-132$^\dagger$         & 1\\
\noalign{\smallskip}\hline
\end{tabular}
\end{table}

\begin{table}
\caption{List of parent nuclides identified in \cite{Nichols2} that should be measured using the total absorption technique to improve the 
predictions of the decay heat in reactors based on $^{233}$U/$^{232}$Th fuel. The list does not contain several relevant cases already measured 
by \cite{Greenwood} and already included in Table \ref{tab:2} (marked with $\dagger$), for more details see \cite{Nichols2}. Rel. (relevance) stands for the priority of the measurement. 
Isotopes marked with asterisks show the measurements performed by our collaboration. For more details in the notation see Table \ref{tab:2}.}
\label{tab:3}       
\begin{tabular}{l c l c l c}
\hline\noalign{\smallskip}
Isotope & Rel. & Isotope & Rel. & Isotope & Rel.\\
\noalign{\smallskip}\hline\noalign{\smallskip}
34-Se-85               & 1 & 38-Sr-92                    & 2 & 51-Sb-128m      & 2\\
34-Se-86               & 2 & 39-Y-96m$^*$          & 1 & 51-Sb-129m      & 2\\
35-Br-84               & 2 & 39-Y-97                     & 1 & 51-Sb-130m      & 1\\
35-Br-89               & 1 & 40-Zr-98                    & 1 &  51-Sb-133         & 2\\
36-Kr-87              & 2 & 41-Nb-99m                & 2 &  54-Xe-138         & 1\\
36-Kr-91              & 1 & 41-Nb-100m$^*$      & 1 &  56-Ba-139         & 2\\
37-Rb-88             & 2 & 41-Nb-102m$^*$      & 1 &  57-La-141         & 2\\
37-Rb-94$^*$     & 1 & 42-Mo-101                & 1 &  57-La-146m      & 2\\ 
\noalign{\smallskip}\hline
\end{tabular}
\end{table}

Several ingredients of this method depend on decay data. The determination of the activities of the fission products ($\lambda_i N_i(t)$) requires a knowledge of the half-lives of the decaying isotopes. The other important quantities are the mean energies released per decay ($\overline{E}_{\beta,i},  \overline{E}_{\gamma,i}$). The mean energies released per decay $i$ can be obtained by direct measurements as in the systematic studies by Rudstam {\it et al.} \cite{Rudstam} and Tengblad {\it  et al.} \cite{Tengblad}. These integral  measurements (energy per decay) require specific setups that are only sensitive to the energy of interest and a careful treatment of all possible systematic errors. Alternatively the mean energies can be deduced from available decay data in nuclear databases such as the Evaluated Nuclear Structure Data File (ENSDF) \cite{Ensdf} if the decay properties are properly known. The term "properly known" beta decay implies a knowledge of the $Q_\beta$ value of the decay, the half-life, the beta distribution probability to the levels in the daughter nucleus and the decay branching ratios of the populated levels. If all this information is available, then it is possible to deduce the mean energies released by the decay using the following relations:

\begin{subequations}
\begin{align}
\overline{E}_{\gamma}  & = \sum_j I_j*E_j ,   \\
\overline{E}_{\beta}  & = \sum_j I_j*<E_\beta>_j ,
\label{eq:mean-energy}
\end{align}
\end{subequations}

where $E_j$ is the energy of the level $j$ in the daughter nucleus, $I_j$ is the probability of a beta transition to level $j$, and $<E_\beta>_j$ is the mean energy of the beta continuum populating level $j$. As can be seen from the formula the mean gamma energy is approximated by the sum of the energy levels populated in the decay weighted by the beta transition probability. This approximation assumes that each populated level decays by gamma deexcitation and ignores conversion electrons which are taken into account in the complete treatment of the mean energy calculations. The mean beta energy, because of the continuum character of the beta distribution emitted in the population of each level, requires the determination of the released mean energy $<E_\beta>_j$ for each 
end-point energy of the beta transition ($Q_\beta-E_j$). Then the mean beta energy ($\overline{E}_{\beta}$)  is obtained as the weighted sum of the mean beta energies populating each level by the beta transition probability. For the determination of $<E_\beta>_j $ for each level one needs to make assumptions about the type of the beta transition (allowed, first forbidden, etc.) and the knowledge of the $Q_\beta$ value of the decay is needed to determine the beta transition end-points. 

Pandemonium can have an impact in the determination of the mean energies from data available in databases. If the beta decay data suffers from the Pandemonium effect the beta decay probability distribution is distorted. This distortion, which implies  increased beta probability to lower lying levels in the daugther nucleus, causes an underestimation of the mean gamma energy and an overestimation of the mean beta energy. This is why TAGS measurements are relevant to this application. 

In fission more than 1000 fission products can be produced. But not all of them are equally important. When addressing a particular problem, like the decay heat, it is of interest to identify which are the most relevant contributors among the large number of fission products. A group of experts working for the International Atomic Energy Agency (IAEA)  \cite{WPEC25} identified high priority lists of nuclei that are important contributors to the decay heat in reactors and that should be measured using the TAGS technique. These lists included nuclides that are produced with high yields in fission and for which the decay data was suspected of suffering from the Pandemonium effect. One argument used for this last selection was if the decay data shows no levels fed in the daughter nucleus in the upper 1/3 excitation energy window of the $Q_\beta$ value.  It is worth noting that this can be considered only as an indication of Pandemonium and not a rigorous rule. Another way of looking for questionable (or odd) data was to look for cases that show different mean energies in the different international databases. The final lists were published in two separate reports, first for U/Pu fuels \cite{Nichols1} and then later for the possible future Th/U fuel \cite{Nichols2}.

In 2004 we started a research programme aimed at studying the beta decay of nuclei making important contributions to the decay heat in reactors. For the planning of any nuclear physics experiment the first step is to decide the best facility to perform it in terms of the availability of the beams, their cleanness and their intensity.  Since some of the most important contributors to the priority list for  $^{235}$U and $^{239}$Pu fission were refractory elements like Tc, Mo and Nb, the options were very limited. In a classical ISOL facility like ISOLDE, the development of a particular beam can take some time if the beam of a particular element is not available.  It is a lengthy and complex task to find the optimum chemical and physical conditions in the ion source for the extraction of a particular element. We decided that the best option concerning the availability of the beams was to perform the measurements at the IGISOL facility in Jyv\"askyl\"a (Finland) \cite{Igisol}. The reason for that was the development of the ion-guide technique. The ion-guide technique, and more specifically the fission ion guide, allows the extraction of fission products independently of the element. In this technique, fission is produced by bombarding  a thin target of natural U with a proton beam. The fission products that fly out of the target are stopped in a gas and transported through a differential pumping system into the first accelerator stage of the mass separator. The dimensions of the ion guide and the pressure conditions are optimized in such a way that the process is fast enough for the ions to survive as singly charged ions. As a result the system is chemically insensitive and very fast (sub-ms) \cite{Ionguide} allowing the extraction of any element including those that are refractory. 

Another important advantage of performing experiments at IGISOL is the availability of the JYFLTRAP Penning trap  \cite{JYFLTRAP} developed for high precision mass measurements at this facility. JYFLTRAP can also be used as a high resolution mass separator for trap assisted spectroscopy measurements, providing a mass resolving power ($\frac{M}{\Delta M}$) of the order of 100 000 to be compared with the resolving power of approximately 500 of the IGISOL separator magnet. The purity of the beams is particularly important for calorimetric measurements like those with TAGS since it reduces systematic errors that can be associated with  contamination of the primary radioactive beams. This advantage has also been used in other types of calorimetric measurements at IGISOL such as the measurements of beta delayed-neutrons using $^{3}$He counters embedded in a polyethylene matrix \cite{Agramunt}.

\begin{figure}
\resizebox{0.50\textwidth}{!}{ \includegraphics{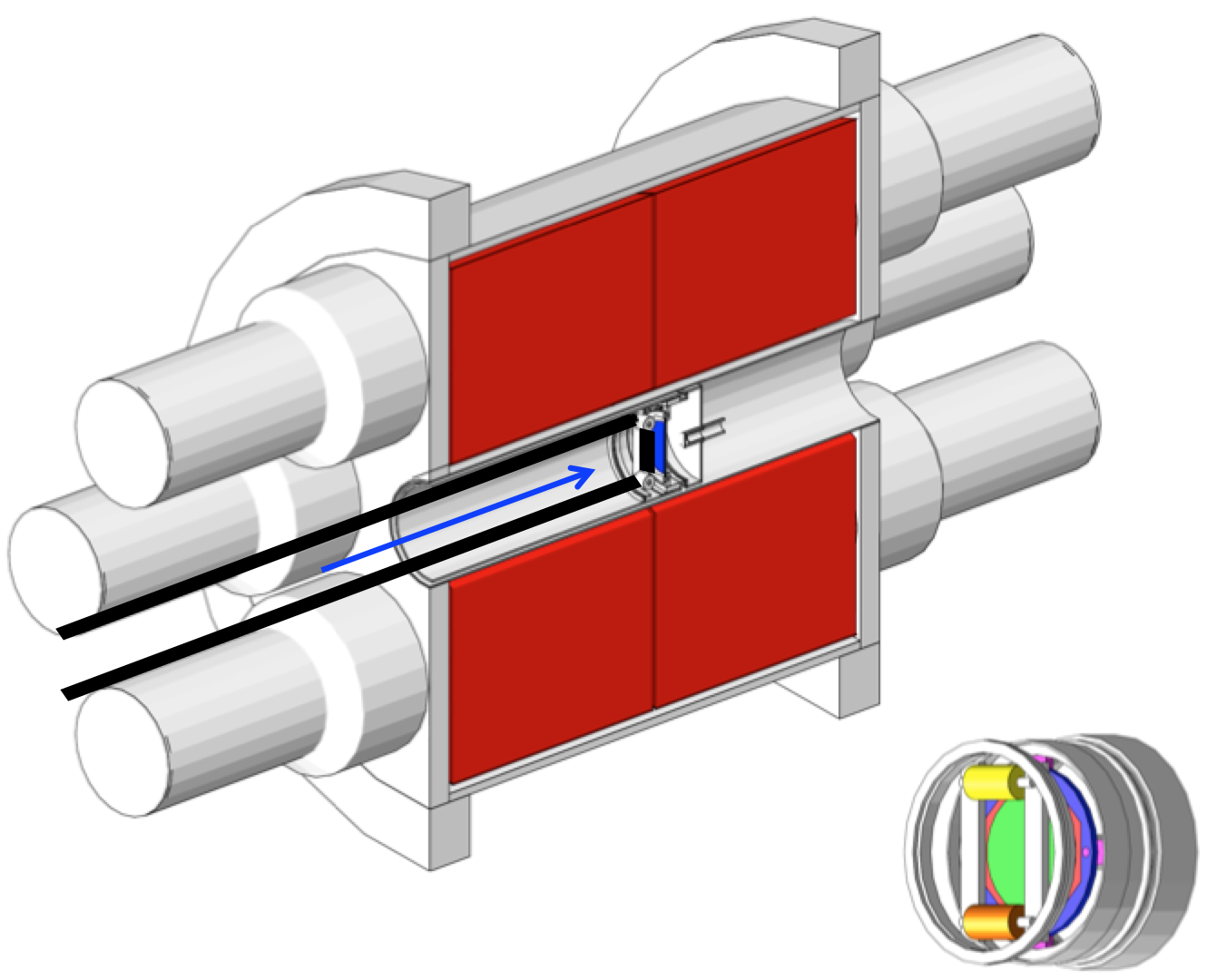}}
\caption{Schematic picture of the {\it Rocinante} total absorption spectrometer used in one of the experiments performed at the IGISOL facility of the University of Jyv\"askyl\"a. The spectrometer is composed of 12 BaF$_2$ crystals. In the lower part the endcap with the Si detector is also presented (not in scale). The thick black lines represent the tape used to move away the remaining activity and the blue line represents the direction of the pure radioactive beam that is implanted in the centre of the spectrometer. Reprinted figure with permission from \cite{Valencia}, Copyright (2017) by the American Physical Society.}
\label{Exp-setup}       
\end{figure}

Three experimental campaigns have been performed at the IGISOL facility to study the beta decay of important contributors to the decay heat and to the antineutrino spectrum in reactors using the TAGS technique \cite{Jyv1,Jyv2,Jyv3,Jyv4}. One of the total absortion setups used in the experiments is presented in Figure \ref{Exp-setup} ({\it Rocinante} TAGS). In a typical experiment, the radioactive beam extracted from IGISOL is first mass separated using the separator magnet and then further separated using the JYFLTRAP Penning trap. Then the beam is transported to the measuring position, at the centre of the total absorption spectrometer where it is implanted in a tape from a tape transport system. The tape is moved in cycles, which are optimized depending on the half-life of the decay of interest. As mentioned earlier, the reason for using a tape transport system is to reduce the effect of undesired daughter, grand-daughter, etc., decay contaminants in the measured spectrum. If necessary, these contaminants have to be substracted from the measured TAGS spectrum and require dedicated measurements. In this kind of measurement the TAGS detector is usually combined with a beta detector as shown in the inset to the Figure \ref{Exp-setup}. The beta detector is used to select coincidences of the beta particles with the TAGS spectrum, which essentially eliminates the effect of the ambient background. 

\begin{figure}
\resizebox{0.50\textwidth}{!}{ \includegraphics{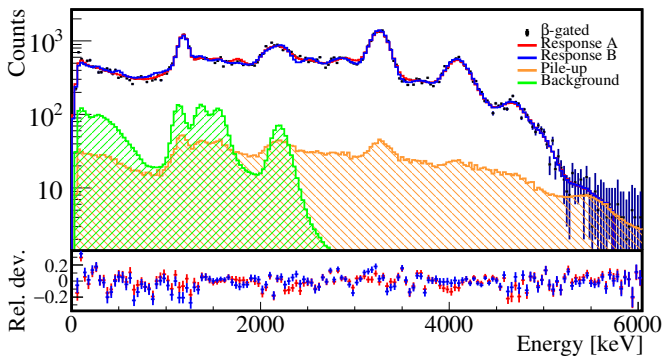}}
\caption{Relevant histograms used in the analysis of the beta decay of $^{86}$Br: measured spectrum (squares with errors), reconstructed spectrum
response A (red line), reconstructed spectrum response B (blue line), summing-pileup contribution (orange line), background (green line). In the lower panel the relative differences of the experimental spectrum vs the reconstructed spectrum are shown. Response A corresponds to the conventional analysis. Response B corresponds to a modified branching ratio matrix to reproduce the measured gamma-ray intensities. For more details see  Rice {\it et al.} \cite{Rice}. Reprinted figure with permission from \cite{Rice}, Copyright (2017) by the American Physical Society.}
\label{86Br_spectrum}       
\end{figure}

\begin{figure}
\resizebox{0.50\textwidth}{!}{ \includegraphics{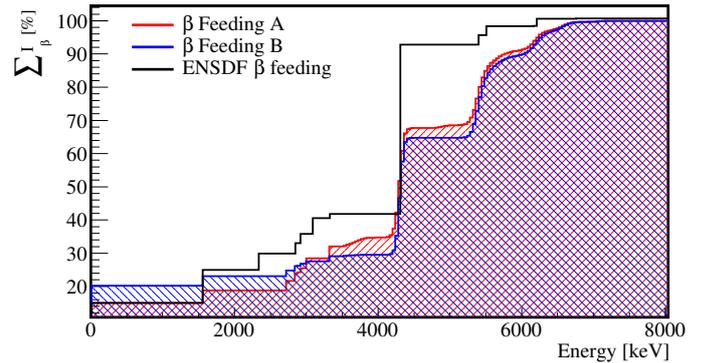}}
\caption{Comparison of the accumulated feeding distributions
obtained in the work of Rice {\it et al.} \cite{Rice} for the decay of $^{86}$Br with available high resolution measurements.  Feeding A and B stands for the TAGS feedings determined. For more details see the text. Reprinted figure with permission from \cite{Rice}, Copyright (2017) by the American Physical Society.}
\label{86Br_feeding}       
\end{figure}

In Figure \ref{86Br_spectrum} we show an example of the recently measured $^{86}$Br decay, which was considered  priority one for the decay heat problem. The spectrum shows the TAGS spectrum obtained with  the beta-gate condition, and the contribution of the different contaminants. The analysis was performed as described earlier. Known levels up to the excitation value of 3560 keV were taken from the compiled high resolution data from \cite{ensdf-86Br}. From 3560 keV to the $Q_\beta$=7633(3) keV value, the statistical model is used to generate a branching ratio matrix using a level density function resulting from a fit to levels from \cite{Goriely-ripl3} and corrected to reproduce the level density at low excitation energy, and E1, M1, and E2 gamma strength functions taken from \cite{Gamma-stf}. For more details see \cite{Rice,Rice-thesis}. In Figure \ref{86Br_spectrum} we also show the comparison of the beta gated TAGS spectrum with the results from the analysis. The reconstructed spectra are obtained by multiplying the response function of the detector with the final feeding distribution obtained from the analysis. In this particular case two results are presented.  Response A corresponds to the conventionally calculated branching ratio matrix that fits better the experimental spectrum. Response B corresponds to a modified branching ratio matrix to reproduce the measured gamma-ray intensities de-exciting each level as measured in high resolution experiments. In Figure \ref{86Br_feeding} the feeding distributions obtained are compared with the available high resolution measurements. This comparison shows that $^{86}$Br decay was suffering from the Pandemonium effect.

In Table \ref{mean_energies} we show a summary of the mean energies deduced from TAS analyses obtained in our measurements performed at Jyv\"askyl\"a. It shows that most of the cases addressed from the high priority list were suffering from the Pandemonium effect. Two cases, that originally were suspected to be Pandemonium cases ($^{102}$Tc, $^{101}$Nb), were not.  Those cases also show the necessity of the TAGS measurements to confirm the suspicion of Pandemonium. Clearly the non existence of feeding in the last $Q_{\beta}/3$ excitation is a good indicator to select cases, but it is not always conclusive. The $^{102}$Tc, $^{101}$Nb cases have strong ground state feedings, which reduces the impact of the undetected gamma branches at high excitation. This is clearly reflected in the differences of the deduced mean energies, when they are compared with the mean energies deduced from high resolution data. The values for one-third of the $Q_{\beta}$-value ($Q_{\beta}/3$) are also given for comparison with the mean energies. The $Q_{\beta}/3$ value is an approach sometimes used by databases, when there is lack of experimental data, which in practical terms divides the available decay energy in equal parts between the mean gamma, beta and antineutrino energies. In the table the mean energies deduced from the high resolution data (ENDSF database) are also given for comparison \cite{Sonzogni_private}.

\begin{table}
\caption{Mean gamma and beta energies deduced from our analyses of beta decays studied at Jyv\"askyl\"a  in comparison with the values deduced from high resolution measurements (ENSDF database). The highest level identified in the decay studies using high resolution and the decay $Q_{\beta}$ values are also given for completeness (for more details see the text). }
\label{mean_energies}       
\resizebox{0.48\textwidth}{!}{
\begin{tabular}{c c c c c c c c}
\hline\noalign{\smallskip}
Isotope & High. Lev. &$Q_{\beta}$  &$Q_{\beta}/3$ & $\overline{E}_\gamma^{HR}$&$\overline{E}_\gamma^{TAGS}$& $\overline{E}_\beta^{HR}$& $\overline{E}_\beta^{TAGS}$\\
\noalign{\smallskip}\hline\noalign{\smallskip}
   $^{86}$Br	&6768    &7633(3)     &2544       &  3360(110)           & 3782(116)   &  1900(300)         & 1687(60)   \\
   $^{87}$Br	&5793    &6818(3)     &2273       &  3100(40)           & 3938($^{+40}_{-67}$)        &   1660(80)        & 1170($^{+32}_{-19}$)       \\
   $^{88}$Br	&6999    &8975(4)     &2992       &  2920(50)           & 4609($^{+78}_{-67}$)        &    2240(240)       & 1706($^{+32}_{-38}$)       \\ 
   $^{91}$Rb	&4793    &5907(9)     &1969       &   2270(40)          & 2669(95)  &     1580(190)      & 1389(44)   \\ 
   $^{92}$Rb	&7363    &8095(6)     &2698       &   170(9)              &  461(14)           &   3640(30)        &  3498(105)          \\    
   $^{94}$Rb	&6064    &10281(8)    &3427       &  1750(50)           & 4063($^{+62}_{-66}$)        &   2020(90)        & 2450($^{+32}_{-30}$)        \\  
   $^{95}$Rb	&4661    &9284(21)   &3095     &  2050(40)  &   3110($^{+17}_{-38}$)      & 2320(110)          & 2573($^{+18}_{-8}$)  \\  
   $^{100gs}$Nb &3130   &6384(21)   &2128   & 710(40)      & 959(318)      & 2540(210) & 2414(154)  \\
   $^{100m}$Nb  &3647  &6698(31)    &2233  & 2210(60)     & 2763(27)      & 2000(200) & 1706(13)  \\ 
   $^{101}$Nb	&1099    &4569(18)    &1523       & 270(22)     & 445(279)    & 1800(300) & 1797(133)  \\
   $^{102gs}$Nb &2480  &7210(40)     &2403  & 2090(100)     & 2764(57)      & 2280(170) & 1948(27)  \\
   $^{102m}$Nb  &1245  &7304(40)    &2435  &                    & 1023(170)    &                  & 2829(82)  \\   
   $^{105}$Mo	&2766    &4953(35)    &1651       & 551(24)     & 2407(93)    & 1900(120) & 1049(44)   \\
   $^{102}$Tc	&2909    &4532(9)     &1511       & 81(4)       & 106(23)     & 1945(16)  & 1935(11)   \\
   $^{104}$Tc	&4268    &5600(50)    &1867       & 1890(30)    & 3229(24)    & 1590(70)  & 931(10)    \\
   $^{105}$Tc   &2403    &3644(35)    &1215       & 671(19)     & 1825(174)   & 1310(210) & 764(81)    \\
   $^{106}$Tc	&3930    &6547(11)    &2182       & 2190(50)    & 3132(70)    & 1900(70)  & 1457(30)   \\
   $^{107}$Tc	&2680    &4820(90)    &1607       & 511(11)     & 1822(450)   & 1890(240) & 1263(212)  \\
   $^{137}$I	&5170   &5880(30)    & 1960      & 1071(2)    & 1220($^{+121}_{-74}$)   & 1897(15) & 1934($^{+35}_{-56}$)  \\
\noalign{\smallskip}\hline
\end{tabular}
}
\end{table}

Our results can also be compared where possible with the results of Greenwood {\it et al.} \cite{Greenwood} and Rudstam {\it et al.} \cite{Rudstam}. Greenwood and co-workers performed a systematic study of fission products at the Idaho National Engineering Laboratory (INEL),  Idaho Falls, USA, using a $^{252}$Cf source and the He-jet technique. They employed a total absorption spectrometer built of NaI(Tl) with the following dimensions,  $ 25.4 \,  \mathrm{cm} \diameter \times 30.5 \,  \mathrm{cm}$ length with a $ 5.1 \, \mathrm{cm} \, \diameter \times 20.3 \, \mathrm{cm} $ long axial well. The analysis technique employed in those studies was based on the creation of a level scheme using the information from high resolution measurements and complementing it with the addition of "pseudo" levels by hand when necessary. The response of the detector was then calculated for the assumed level scheme using the Monte Carlo code  CYLTRAN  (for more details see \cite{Greenwood_MC}). The systematic study of Greenwood {\it et al.} provided TAGS data for 48 decays including the decay of three isomeric states. Since a different analysis technique was used, it is interesting to compare the results of Greenwood with the more recent results and look for possible systematic deviations for cases where the comparison is possible.  



As mentioned earlier Rudstam {\it et al.} performed systematic measurements of gamma and beta spectra and deduced mean energies \cite{Rudstam,Tengblad} of interest for the prediction of decay heat. Beta spectra of interest for the prediction of the antineutrino spectrum from reactors were also measured \cite{Tengblad}. The direct measurements were performed at ISOLDE and at the OSIRIS separator using setups optimized for the detection of the gamma- and beta-rays emitted by the fission products. In the case of the mean gamma energies, the setup required an absolute gamma efficiency calibration with the assumption that the decay used in the calibration did not suffer from the Pandemonium effect \cite{Rudstam}. In their measurements the beta decay of $^{91}$Rb was used as a calibration. This decay with a $Q_\beta$ of 5907(9) keV shows a complex decay scheme from high resolution measurements that populates levels up to 4700 keV in the daughter nucleus.  So it was assumed naturally by the authors of \cite{Rudstam} that this decay was not a Pandemonium case (in the publication \cite{Rudstam} it is mentioned that the gamma spectrum extends up to 4500 keV). 

\begin{figure}
\resizebox{0.48\textwidth}{!}{ \includegraphics{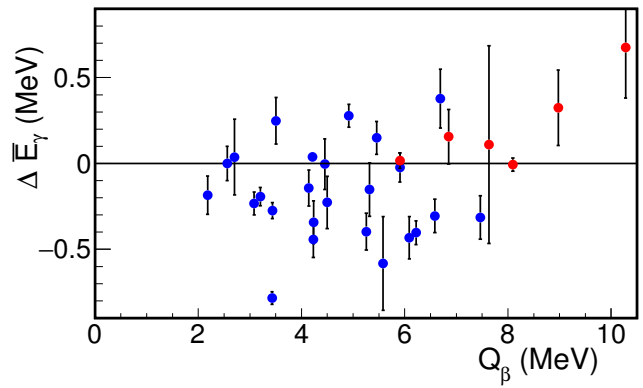}}
\caption{Differences between the mean gamma energies obtained with TAGS measurements (see Table \ref{comparison_rudstam} and the text for more details) and the direct measurements of Rudstam {\it et al.} \cite{Rudstam} after renormalization by a factor of 1.14, which was deduced from the comparison of our newly determined mean energy for the decay of  $^{91}$Rb \cite{Rice} with that employed by Rudstam {\it et al.} \cite{Rudstam}. Blue points represent Greenwood TAGS data and red points represent results from our collaboration. A systematic difference with a mean value of -180 keV remains. Reprinted figure with permission from \cite{Rice}, Copyright (2017) by the American Physical Society.}
\label{comparison_differences}       
\end{figure}

In a contribution to the Working Party on International Evaluation Co-operation of the NEA Nuclear Science Committee (WPEC 25)
group activities \cite{WPEC25}, the late O. Bersillon performed a comparison between the Greenwood and Rudstam mean gamma and beta energies for those decays that were available in both data sets. This comparison was revisited in \cite{Tain-preprint}. A clear systematic difference was shown, pointing to possible systematic errors in one or in both data sets \cite{Valencia,Rice,Tain-preprint}. For that reason it was  decided to measure the $^{91}$Rb decay using the TAGS  technique, to see if the decay was suffering from the Pandemonium effect or not and accordingly check if this decay was adequate as an absolute calibration point to obtain mean gamma energies in \cite{Rudstam}. 
The 2669(95) keV mean gamma value obtained from our measurements \cite{Rice} can be compared with the value used by Rudstam {\it et al.}  (2335(33) keV) showing that this decay suffered from Pandemonium and also showing the necessity of renormalizing the Rudstam data by a factor of 1.14. With this renormalization, the mean value of the differences between the two data sets (TAGS vs Rudstam) reduces from -360 keV to -180 keV, but still the discrepancy remains \cite{Rice}.  This is shown in Figure \ref{comparison_differences}  and Table \ref{comparison_rudstam}. It should be noted that our mean gamma energy value for this decay agrees nicely with the value obtained by Greenwood {\it et al.} (2708(76) keV) \cite{Greenwood}. 

\begin{table}
\caption{Comparison of mean gamma energies obtained with the TAGS measurements ($\overline{E}_{\gamma}^{T}$) with those obtained in the dedicated measurements by Rudstam {\it et al.} ($\overline{E}_{\gamma}^{R}$) \cite{Rudstam} (original values, not renormalized).  Marked with asterisks are our TAGS results for the mean energies. The TAGS results not marked with asterisks are taken from Greenwood {\it et al.}  \cite{Greenwood}. 
}
\label{comparison_rudstam}       
\resizebox{0.48\textwidth}{!}{
\begin{tabular}{c c c c c c}
\hline\noalign{\smallskip}
Isot. & $\overline{E}_{\gamma}^{T}$&$\overline{E}_{\gamma}^{R}$&Isot. & $\overline{E}_{\gamma}^{T}$&$\overline{E}_{\gamma}^{R}$\\
\noalign{\smallskip}\hline\noalign{\smallskip}
$^{86}$Br$^*$      & 3782(116)  & 3420(500)  &     $^{95}$Y         & 1223(50)  & 1060(120)   \\
$^{87}$Br$^*$      & 3938($^{+40}_{-67}$)  & 3560(130)  &    $^{138m}$Cs      & 426(27)   & 500(80)     \\
$^{88}$Br$^*$    & 4609($^{+78}_{-67}$)   & 4290(180)  &	      $^{139}$Cs       & 305(8)	& 299(21)     \\
$^{89}$Rb          & 2228(145) & 1740(40)   &       $^{140}$Cs       & 1864(37)  & 1270(50)    \\
$^{90}$Rb          & 2272(79)  & 1710(50)   &        $^{141}$Cs       & 1708(29)  & 1140(90)    \\
$^{90m}$Rb       & 3866(115) & 3690(110)  &      $^{141}$Ba       & 906(27)   & 620(40)     \\
$^{91}$Rb          & 2708(76)  & 2335(33)   &         $^{142}$Ba       & 1059(64)  & 760(80)     \\
$^{91}$Rb$^*$      & 2669(95)  & 2335(33)   &	$^{143}$Ba       & 1343(49)  & 870(100)    \\
$^{92}$Rb$^*$       & 461(14)   & 393(32)    &	$^{144}$Ba       & 785(33)   & 480(50)     \\
$^{93}$Rb       & 2523(53)  & 1920(100)  &            $^{145}$Ba       & 1831(44)  & 1460(130)   \\
$^{93}$Rb$^*$       & 2397(25)  & 1920(100)  &	$^{143}$La       & 424(9)	& 130(40)     \\
$^{94}$Rb$^*$       & 4063($^{+62}_{-66}$)  & 4120(250)  &	$^{144}$La       & 3158(68)  & 2240(230)   \\
$^{93}$Sr       & 2167(68)  & 1760(70)   &              $^{145}$La       & 2144(52)  & 1480(80)    \\
$^{94}$Sr       & 1419(135) & 1450(10)   &              $^{145}$Ce       & 885(59)   & 770(70)     \\
$^{95}$Sr       & 1790(43)  & 1180(100)  &              $^{147}$Ce       & 1497(35)  & 620(10)     \\
$^{94}$Y        & 757(34)   & 900(50)    & 	         $^{147}$Pr       & 929(32)   & 840(190)    \\

\noalign{\smallskip}\hline
\end{tabular}
}
\end{table}

\begin{figure}
\resizebox{0.53\textwidth}{!}{ \includegraphics{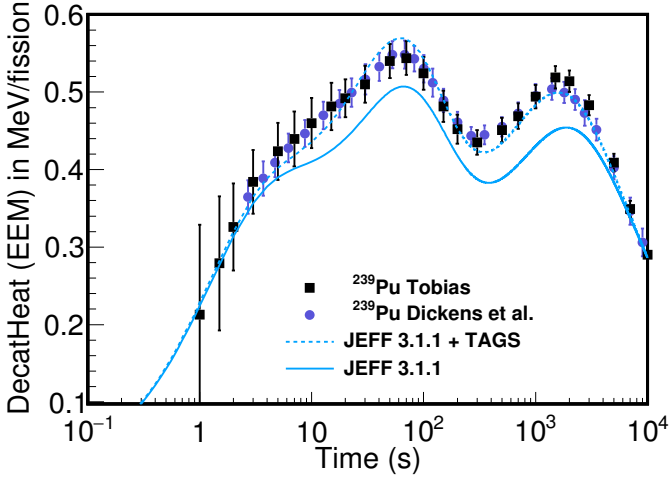}}
\caption{Impact of the inclusion of the total absorption measurements performed 
for 13 decays ($^{86,87,88}$Br, $^{91,91,94}$Rb, $^{101}$Nb, $^{105}$Mo, $^{102,104,105,106,107}$Tc) published in Refs. \cite{Algora,Jordan,Valencia,Rice,Zak} in the gamma component of the decay heat calculations for $^{239}$Pu.}
\label{DH_239Pu_1}       
\end{figure}

\begin{figure}
\resizebox{0.53\textwidth}{!}{ \includegraphics{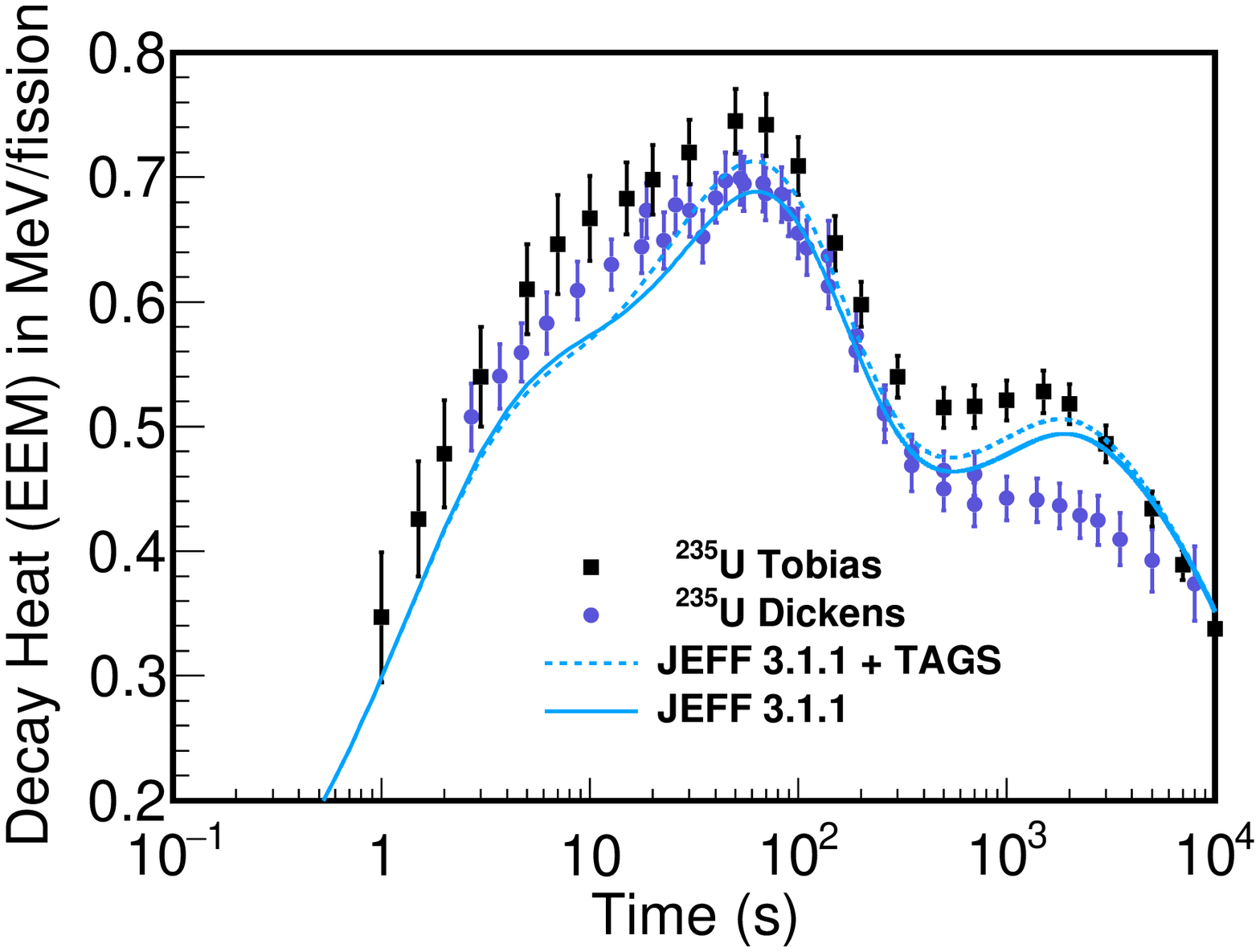}}
\caption{Impact of the inclusion of the total absorption measurements performed for 13 decays in the gamma component of the decay heat calculations for $^{235}$U (see Figure \ref{DH_239Pu_1} for more details).}
\label{DH_235U_1}       
\end{figure}

\begin{figure}
\resizebox{0.53\textwidth}{!}{ \includegraphics{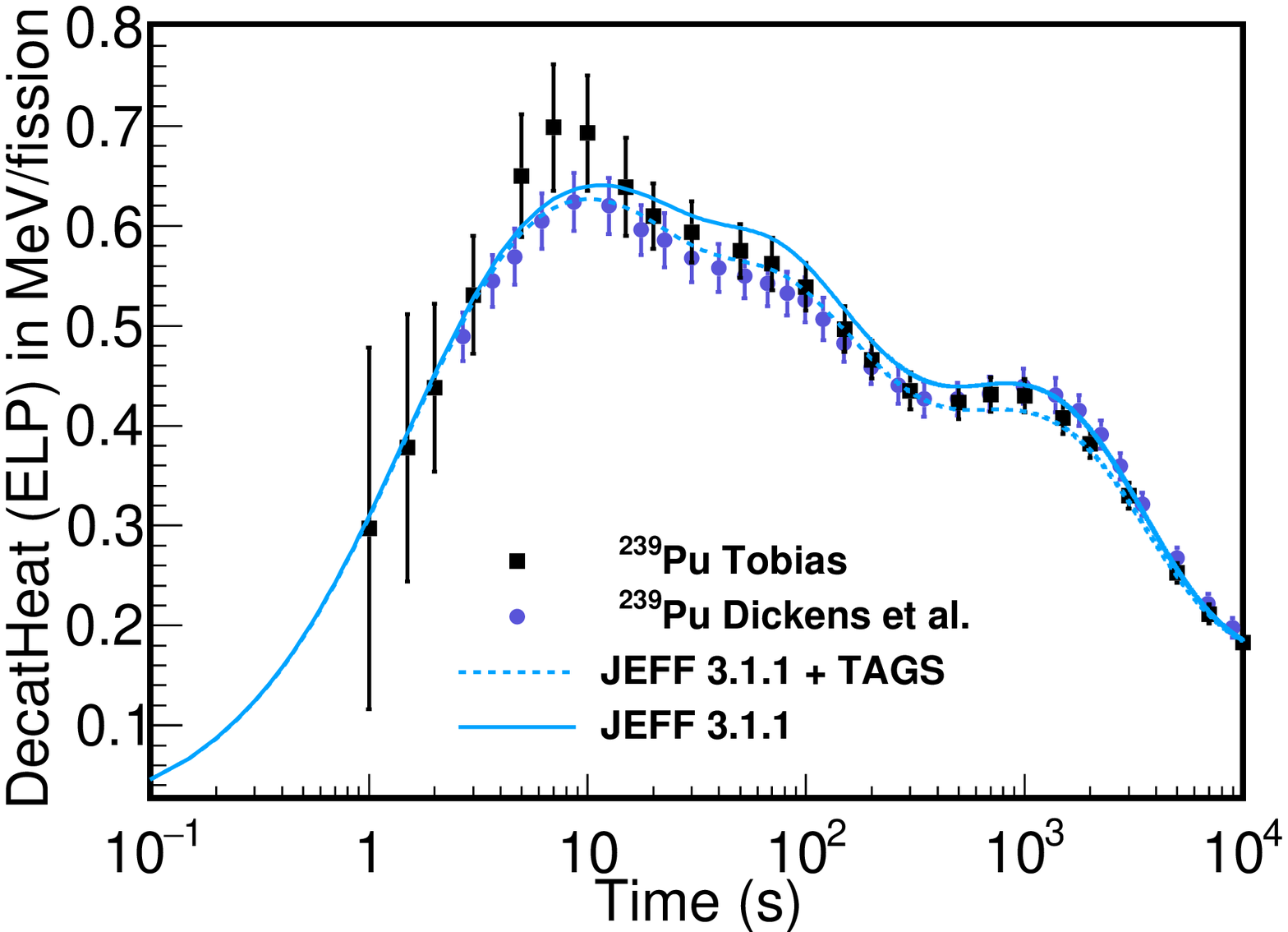}}
\caption{Impact of the inclusion of the total absorption measurements performed for 13 decays  in the beta component of the decay heat calculations for $^{239}$Pu (see Figure \ref{DH_239Pu_1} for more details).}
\label{DH_239Pu_2}       
\end{figure}

\begin{figure}
\resizebox{0.53\textwidth}{!}{ \includegraphics{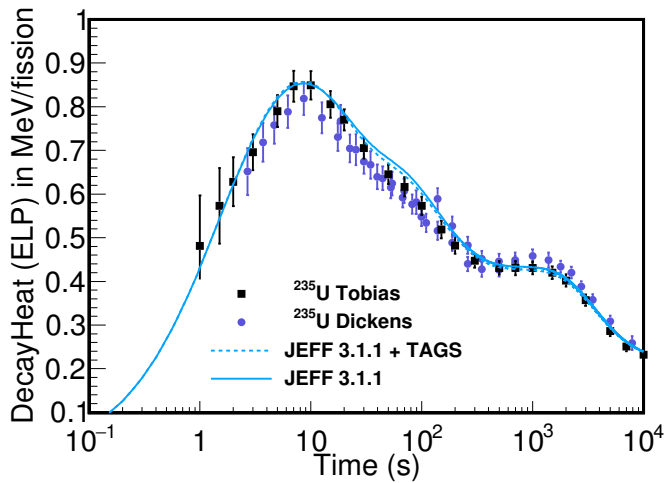}}
\caption{Impact of the inclusion of the total absorption measurements performed for 13 decays  in the beta component of the decay heat calculations for $^{235}$U (see Figure \ref{DH_239Pu_1} for more details).}
\label{DH_235U_2}       
\end{figure}

In Figs. \ref{DH_239Pu_1}($^{239}$Pu) and \ref{DH_235U_1} ($^{235}$U) the total impact of the total absorption measurements of the 13 decays ($^{86,87,88}$Br, $^{91,92,94}$Rb, $^{101}$Nb, $^{105}$Mo, $^{102,104,105,106,107}$Tc) published in Refs. \cite{Algora,Jordan,Valencia,Rice,Zak} is presented in comparison with the decay heat measurements reported by Tobias \cite{Tobias} and Dickens \cite{Dickens} for the electromagnetic component ($E_\gamma$) of the decay heat. Similarly in Figs. \ref{DH_239Pu_2}-\ref{DH_235U_2} the impact of the same decays is compared for the charged particle component ($E_\beta$) of the decay heat. To show the impact of the total absorption measurements the data base JEFF3.1.1 is used, in which no total absorption data is included. Calculations are performed using the bare JEFF 3.1.1 and a modified version of the JEFF3.1.1 database with the inclusion of the total absorption data for the mean energies. The results presented were provided by Dr. L. Giot \cite{Giot}. In the figures, it can be noted the large impact of the mentioned decays and the relevance of the total absorption measurements for a proper assesment of the decay heat based on summation calculations.

From the Figure \ref{DH_235U_1}, it is clear that additional measurements are needed for improving the description of the $^{235}$U fuel, and new measurements are certainly required for future fuels like the $^{233}$U/$^{232}$Th case. 


\section{Neutrino applications}
\label{neutrino}
Nuclear reactors constitute an intense source of electron antineutrinos, with typically $10^{20}$ antineutrinos per second emitted by a 1GWe reactor. The reactor at Savannah River was the site of the discovery of the neutrino  in 1956 by Reines and Cowan \cite{Reines}, thus confirming Pauli's predictions of twenty-six years earlier \cite{Pauli}. Just like the decay heat described above, antineutrinos arise from the beta decays of the fission products in-core. Their energy spectrum and flux depend on the distribution of the fission products which reflects the fuel content of a nuclear reactor. This property combined with the fact that neutrinos are sensitive only to the weak interaction could make antineutrino detection a new reactor monitoring tool~\cite{Mikaelian}. Both particle and applied physics are the motivations of their study at power or research reactors nowadays with detectors of various sizes and designs placed at short or long distances.  
In the last decade, three large neutrino experiments with near and far detectors, were installed at Pressurized Water Reactors ~\cite{DC,DayaBay,RENO}, to try to pin down the value of the $\theta_{13}$ mixing angle parameter governing neutrino oscillations. These experiments have sought the disappearance of antineutrinos by comparing the flux and spectra measured at the two sites, both distances being carefully chosen to maximise the oscillation probability at the far site.
The three experiments ~\cite{DC,DayaBay,RENO} have now achieved a precise measurement of the $\theta_{13}$ mixing angle, paving the way for future experiments at reactors looking at the neutrino mass hierarchy or for experiments at accelerators for the determination of the delta phase, that governs the violation of the CP symmetry in the leptonic sector, thus shedding light on why there is an abundance of matter rather than antimatter in the Universe.

Though they have used one or several near detectors in order to measure the initial flux and energy spectrum of the emitted antineutrinos, the prediction of the latter quantities still enters in the systematic uncertainties of their measurements, because their detectors are usually not placed on the isoflux lines of the several reactors of the plant at which they are installed~\cite{Cucoanes2013}. In addition, the Double Chooz experiment started to take data with the far detector alone, implying the need to compare the first data with a prediction of the antineutrino emission by the two reactors of the Chooz plant.
Two methods employed to calculate reactor antineutrino energy spectra were revisited at that time, i.e. the conversion and the summation methods.

The conversion method consists in converting the integral electron spectra measured at the research reactor at the Institute  Laue-Langevin (ILL) in Grenoble (France) by Schreckenbach and co-workers~\cite{SchreckU5-1,SchreckU5-2,SchreckU5Pu9,Hahn} with $^{235}$U, $^{239}$Pu and $^{241}$Pu thin targets under a thermal neutron flux. These spectra exhibit rather small uncertainties and remain a reference as no other comparable measurement has been performed since. Being integral measures, no information is available on the individual beta decay branches of the fission products. This prevents the use of the conservation of energy to convert the beta into antineutrino spectra. Schreckenbach {\it et al.} developed a conversion model, in which they used 30 fictitious beta branches spread over the beta energy spectrum to convert their measurements into antineutrinos. In 2011 Mueller {\it et al.}~\cite{Mueller} revisited the conversion method and improved it through the use of more realistic end-points and Z distributions of the fission products, available thanks to the wealth of nuclear data accumulated over 30 years, and through the application of the corrections to the Fermi theory at branch level in the calculation of the beta and antineutrino spectra. After these revisions, the prediction of detected antineutrino flux at reactors compared with the measurements made at existing short baseline neutrino experiments revealed a deficit of 3\%. The result was confirmed immediately by Huber~\cite{Huber} who carried out a similar calculation though he did not explicitly use beta branches from nuclear data. This antineutrino deficit was even increased by the revision of the neutron lifetime and the influence of the long-lived fission products recalculated at the time, to finally amount to 6\%~\cite{Mention}. A new neutrino anomaly was born: the reactor anomaly. Several research leads were followed since to explain this deficit. An exciting possibility is the oscillation of reactor antineutrinos into sterile neutrinos ~\cite{Mention}, which has triggered several new experimental projects worldwide~\cite{SoLid,STEREO,Prospect}. 
In 2015, the mystery deepened when Daya Bay in China, Double Chooz in France and RENO in Korea, reported the detection of a distortion (colloquially called “bump”) in the measured antineutrino energy spectrum with respect to the converted spectrum, which could not be explained by any neutrino oscillation. The three experiments rely on the same detection technique and similar detector designs, which make it possible that they would all suffer from the same detection bias as suggested in \cite{Mention2017}. But the three collaborations have thoroughly investigated this hypothesis without success. In the face of the observed discrepancies between the converted spectra and the measured reactor antineutrino spectra, it is worth considering more closely the existing methods used to compute them and other possible explanations like the case of nonlinear corrections discussed in Section \ref{section_TAS} in relation to the $^{100}$Tc decay \cite{Huber2}.


Converted spectra rely on the unique measurements performed at the high flux ILL research reactor with the high resolution magnetic spectrometer BILL~\cite{BILL}, using thin actinide target foils exposed to a thermal neutron flux that was well under control. This device was exceptional as it allowed the measurement of electron spectra ranging from 2 to 8 MeV in 50~keV bins (smoothed over 250~keV in the original publications) with an uncertainty dominated by the absolute normalization uncertainty of 3\% at 90\% C.L. except for the highest energy bins with poor statistics~\cite{SchreckU5-1,SchreckU5-2,SchreckU5Pu9,Hahn}. 
The calibration of the spectrometer was performed with conversion electron sources or (n,e$^-$) reactions on targets of $^{207}$Pb, $^{197}$Au, $^{113}$Cd and using the beta decay of $^{116}$In
 providing calibration points up to 7.37~MeV. The irradiation duration ranged from 12 hours to 2 days. 
Two measurements of the $^{235}$U electron spectrum were performed, the first one lasting 1.5 days and the second one 12 hours. 
The normalisation of the two spectra disagree because they were normalized using two different (n,e$^-$) reactions on $^{197}$Au and $^{207}$Pb respectively in chronological order.
The measurement retained by the neutrino community is the second one. 
The conversion procedure consists in successive fits of the electron total spectrum with beta branches starting with the largest end-points. The total electron spectrum is fitted iteratively bin by bin starting with the highest energy bins, and the contributions to the fitted bin are subtracted from the total spectrum. The reformulation of the finite size corrections, as well as a more realistic charge distribution of the fission products and a much larger set of beta branches have been the key for the newly obtained converted antineutrino spectra of~\cite{Mueller} and~\cite{Huber}. 
But the possibility remains that the electron and/or converted spectra suffer from unforeseen additional uncertainties. 
Indeed the normalisation of the electron spectra 
relies on the (n,e$^-$) reactions quoted above and on internal conversion coefficient values that may have  both been re-evaluated since ~\cite{OnillonAAP}. In addition, the exact position of the irradiation experiment in the reactor is not well known and may have an impact on the results as well ~\cite{OnillonAAP}.
Another concern is associated with the conversion model itself, where uncertainties may not take into account missing underlying nuclear physics. In Mueller's conversion model, forbidden non-unique transitions are replaced by forbidden unique transitions (when the spins and parities are known!). The shapes of the associated beta and antineutrino spectra are not well known and the forbidden transitions dominate the flux and the spectrum above 4~MeV. Several theoretical works have attempted to estimate the uncertainties introduced by this lack of knowledge~\cite{Hayes,FangBrown,Hayen}. The latest study~\cite{Hayen} reports a potential effect compatible with the observed shape and flux anomalies. Another source of uncertainties comes from the weak magnetism correction entering in the spectral calculation \cite{Huber,Hayes2017} that is not well constrained experimentally in the mass region of the fission products. These two extra uncertainties affect converted spectra and are not included in the published uncertainties. Eventually the conversion process itself could be discussed, as the iterative fitting procedure is not the only possible conversion method and it is suspected of inducing additional uncertainties~\cite{HayesSolvay2017}.

In order to identify what could be at the origin of these anomalies, the understanding of the underlying nuclear physics ingredients is mandatory. Indeed, only the decomposition of the reactor antineutrino spectra into their individual contributions and the study of the missing underlying nuclear physics will allow us to understand fully the problem and provide reliable predictions. The best tool to address these questions is to use the summation method. This method is based on the use of nuclear data combined in a sum of all the individual contributions of the beta branches of the fission products, weighted by the amounts of the fissioning nuclei. Two types of datasets are thus involved in the calculation: fission product decay data, and fission yields. This method was originally developed by~\cite{King} followed by~\cite{Avignonne} and then by~\cite{Tengblad,Vogel81}. The $\beta$/$\bar{\nu}$ spectrum per fission of a fissile isotope $S_k(E)$ can be broken down into the sum of all fission product $\beta$/$\bar{\nu}$ spectra weighted by their activity $\lambda_{i} N_{i}(t)$ similarly to what is done for decay heat calculations:
\begin{equation}
S_k(E)=\sum_{i}{\lambda_i N_i(t)\times S_{i}(E)}
\label{Sk}
\end{equation} 

Eventually, the $\beta$/$\bar{\nu}$ spectrum of one fission product ($S_i$) is the sum over the $\beta$ branches (or beta transition probabilities) of all $\beta$ decay spectra (or associated $\bar{\nu}$ spectra), $S_{i}^{b}$ (in equation~\ref{Sfp}), of the parent nucleus to the daughter nucleus weighted by their respective beta branching ratios according to:
\begin{equation}
S_{i}(E)=\sum_{b=1}^{N_{b}}{f_{i}^{b}\times S_{i}^{b}(Z_{i},A_{i},E_{0 i}^{b},E)}
\label{Sfp}
\end{equation}
 where $f_{i}^{b}$ represents the beta transition probability of the b branch, $Z_{i}$ and $A_{i}$ the atomic number and the mass number of the daughter nucleus respectively and $E_{0 i}^{b}$ is the endpoint of the beta transition b. 
In 1989 the measurement of 111 beta spectra from fission products by Tengblad {\it et al.} ~\cite{Tengblad} was used for a new calculation of the antineutrino energy spectra through the summation method. But the overall agreement with the integral beta spectra measured by Hahn {\it et al.} ~\cite{Hahn} was at the level of 15-20\% showing that a large amount of data were missing at that time. Lately, the summation calculations were re-investigated using updated nuclear databases. Indeed the summation method is the only one able to predict antineutrino spectra for which no integral beta measurement has been performed. The existing aggregate beta spectra needed to apply the conversion method are relatively few and were measured under irradiation conditions that are not exactly the same as those existing in power reactors. Among the discrepancies, the energy distributions of the neutrons generating the fissions in the ILL experiments are different from those in actual power reactors, and even more from the ones in innovative reactor designs such as fast breeder reactors.  The aggregate beta spectra were measured for finite irradiation times much shorter than the typical times encountered in power reactors. These few spectra and the specific conditions are not usable for innovative reactor fuels or require corrections for longer irradiation times (called off-equilibrium corrections) or more complex neutron energy distributions in-core. Until the recent measurement of the $^{238}$U beta spectrum at Garching by~\cite{Haag}, the conversion method could not be applied to obtain a prediction of the $^{238}$U fast fission antineutrino spectrum. 
This was one of the motivations for the first re-evaluation of the summation spectra that was performed in Mueller {\it et al.}, the second being to provide off-equilibrium corrections~\cite{Mueller} to the converted spectra. 
In this work, several important conclusions were already listed regarding summation calculations for antineutrinos. 

The evaluated nuclear databases do not contain enough decay data to supply detailed beta decay properties for all the fission products stored in the fission yields databases. The evaluated databases have thus to be supplemented by other data or by model calculations for the most exotic nuclei. The relative ratio of the aggregate beta spectra with the obtained summation spectra from databases exhibited a shape typical of the Pandemonium effect, with an overestimate of the high energy part of the spectra in the nuclear data. The maximum amount of data free of the Pandemonium effect should thus be included in the summation calculations. The difficulty comes from the fact that these Pandemonium-free data are usually not included in the evaluated databases. One has thus to gather the existing decay data and compute the associated antineutrino spectra. The Pandemonium-free data are mostly existing TAGS measurements~\cite{Greenwood} and the electron spectra directly measured by Tengblad {\it et al.} ~\cite{Tengblad}. They were included in an updated summation calculation performed in~\cite{Fallot}, in which the seven isotopes measured with the TAGS technique that had so much impact on the $^{239}$Pu electromagnetic decay heat, i.e. $^{105}$Mo, $^{102,104-107}$Tc, and $^{101}$Nb~\cite{Algora}, were taken into account. The calculation revealed that these TAGS results had a very large impact on the calculated antineutrino energy spectra, reaching 8\% in the Pu isotopes at 6~MeV. But it appeared that summation calculations still overestimate the beta spectra at high energy, indicating that there were large contributions from nuclei where the data suffer from the Pandemonium effect in the decay databases. The situation is thus similar to that already encountered in the decay heat summation calculations.  These conclusions reinforced the necessity for new experimental TAGS campaigns and spread the message worldwide.
New summation calculations were developed and other experimental campaigns were launched using the TAGS technique~\cite{Sonzo2015,Rasco_let,Alexa}.
In~\cite{Sonzo2015} a careful study of the existing evaluated fission yield databases was performed. It appeared that the choice of the fission yield database had a large impact on the summation spectra obtained, because of  mistakes identified in the ENDF/B-VII.1 fission yields for which corrections were proposed. Once corrected, the ENDF/B-VII.1 fission yields provide spectral shapes in close agreement with the JEFF3.1 fission yields.
In 2012, the  agreement obtained was at the level of 10\% with respect to the integral beta spectra measured at ILL and the number of nuclei requiring new TAGS measurements was considered as achievable.
Lists of priority for new TAGS measurements were established first by the Nantes group~\cite{Zak} (which triggered our first experimental campaign devoted to reactor antineutrinos in 2009), then by the BNL team~\cite{Sonzo2015} and eventually a table based on the Nantes summation method was published in the frame of TAGS consultant meetings organized by the Nuclear Data Section of the IAEA~\cite{Report-NDS-676}.
A portion of the table from~\cite{Report-NDS-676} is shown in Table \ref{Table_neutrino}, with the measurements performed by our collaboration marked with asterisks. More than half of the first priority nuclei have been measured by our collaboration with the TAGS technique. The Oak Ridge group is involved in similar studies, see for example 
the results for several isotopes published in \cite{Rasco_let,Alexa}. 




\begin{table}
\caption{List of nuclides identified by the IAEA TAGS Consultants that should be measured using the total absorption technique to improve the 
predictions of the reactor antineutrino spectra. These nuclides are of relevance for conventional reactors based on $^{235}$U and $^{239}$Pu nuclear fuels. The list contains 34 nuclides
\cite{Report-NDS-676}. Relevance (Rel.) stands for the priority of the measurement. Isotopes marked with asterisks show the measurements performed by our collaboration,  
m stands for metastable or isomeric state.}
\label{Table_neutrino}       
\begin{tabular}{l c l c l c}
\hline\noalign{\smallskip}
Isotope & Rel. & Isotope & Rel. & Isotope & Rel.\\
\noalign{\smallskip}\hline\noalign{\smallskip}
36-Kr-91  	& 2  & 39-Y-97m          & 1 & 53-I-138$^*$     & 2\\
37-Rb-88      & 1 & 39-Y-98m  & 1 & 54-Xe-139       & 1\\
37-Rb-90      & 1 & 39-Y-99$^*$  & 1  & 54-Xe-141   & 2\\
37-Rb-92$^*$            & 1 & 40-Zr-101 & 1 & 55-Cs-139 & 1\\
37-Rb-93$^*$            & 1 & 41-Nb-98$^*$  & 1 & 55-Cs-140$^*$      & 1\\
37-Rb-94$^*$                         & 2 & 41-Nb-100$^*$                    & 1 & 55-Cs-141    & 2\\
38-Sr-95$^*$     & 1 & 41-Nb-101$^*$  & 1 &55-Cs-142$^*$     & 1\\
38-Sr-96                           & 1 & 41-Nb-102$^*$  &1  & 57-La-146                        & 2\\
38-Sr-97                                & 2 & 41-Nb-104m  & 2 &                  & \\
39-Y-94              & 1 & 52-Te-135                     & 1 &                      & \\
39-Y-95$^*$             & 1 & 53-I-136                     & 2 &                       & \\
39-Y-96$^*$            & 1 & 53-I-136m                    & 1\\
39-Y-97    & 2 &  53-I-137$^*$         & 1\\
\noalign{\smallskip}\hline
\end{tabular}
\end{table}

In 2009, $^{91-94}$Rb and $^{86-88}$Br were measured with the {\it Rocinante} TAGS (for details see ~\cite{Valencia,Rice} and Figure \ref{Exp-setup}) placed after the JYFLTRAP Penning trap of the IGISOL facility~\cite{Igisol}. Only $^{92,93}$Rb were in the top ten of the nuclei contributing significantly to the reactor antineutrino spectrum.
$^{92}$Rb itself contributes 16\% of the antineutrino spectrum emitted by a pressurized water reactor (PWR) between 5 and 8~MeV. Its contributions to the $^{235}$U and $^{239}$Pu antineutrino spectra are 32\% and 25.7\% in the 6 to 7~MeV bin and 34 and 33\% in the 7 to 8~MeV bin. In 2009, the ground state (GS) to GS beta intensity of this decay was set to 56\% in ENSDF.  This value was revised in 2014 to 95.2\%. A maximum of 87\%$\pm$2.5\% was deduced from our TAGS data, having a large impact on the antineutrino spectra~\cite{Zak}. This value was confirmed by the Oak Ridge measurements \cite{Rasco_let}. The $^{92}$Rb case is worth noting because it is not a case suffering from Pandemonium, but its GS to GS beta branch was underestimated in former evaluations. In the analysis of this nucleus, the sensitivity of the reconstructed spectrum (and thus of the $\chi^2$ obtained ) to the value of the GS-GS branch was very high, because of the large penetration of the electrons in the TAGS. The quoted uncertainties were obtained by varying the input parameters entering into the analysis, such as the calibration parameters, the thickness of the beta detector, the level density, the normalisation of the backgrounds, etc..
The beta decay data for $^{92}$Rb that were used in the previous summation calculations~\cite{Fallot} were those from Tengblad {\it et al.} The impact of replacing these data with the new TAGS results amounts to 
4.5\% for $^{235}$U, 3.5\% for $^{239}$Pu, 2\% for $^{241}$Pu, and 1.5\% for $ ^{238}$U. A similar impact was found on the summation model developed by Sonzogni {\it et al.}~\cite{Sonzo2015} but a much larger impact (more than 25\% in $^{235}$U) was found on another model in which no Pandemonium-free data were included~\cite{Dwyer}.  

Though not motivated by neutrino physics, $^{86-88}$Br and $^{91, 94}$Rb were measured in the same experiment. $^{86-88}$Br and $^{91}$Rb were not in the priority list of~\cite{Report-NDS-676}. The main motivation for those cases was the study of moderate beta delayed neutron emitters using complementary techniques and the study of the decay used as normalization in the measurements by Rudstam {\it et al} \cite{Rudstam} already mentioned in Section \ref{decay_heat}. These TAGS measurements confirmed the Pandemonium problem in the existing data. $^{86, 87}$Br and $^{91}$Rb did not show a large impact on the antineutrino spectra~\cite{Tain,Valencia,Rice}. 
On the other hand, $^{94}$Rb, ranked as priority 2 in the table from~\cite{Report-NDS-676}, and $^{88}$Br exhibited a quite large impact on the spectra. This was verified with two different summation calculations~\cite{Valencia}. In one of the models, the new TAGS data replaced high resolution spectroscopy data, and thus the observed impact was typical of a correction of the Pandemonium effect i.e. a decrease of the high energy part of the aggregate antineutrino energy spectrum. The impact reached 4\% in $^{235}$U and $^{239}$Pu in the case of $^{94}$Rb and was more modest as regards $^{88}$Br with a 2-3\% decrease in the latter actinides between 8 and 9~MeV. The latter range explains the reason why $^{88}$Br did not belong to the priority list of~\cite{Report-NDS-676} that was established for a contribution to the PWR antineutrino spectrum larger than 1\% between 3 and 8~MeV. In the second model the TAGS data replaced data measured by Tengblad {\it et al.} that were considered in the model because they were assumed to be Pandemonium-free. The replacement of $^{87}$Br had little impact, that of $^{94}$Rb led to a 3\% decrease at 8~MeV but that of $^{88}$Br brought a 7\% increase between 8 and 9~MeV, with a cancellation of the last two effects below 8~MeV.

\begin{figure}
\resizebox{0.48\textwidth}{!}{ \includegraphics{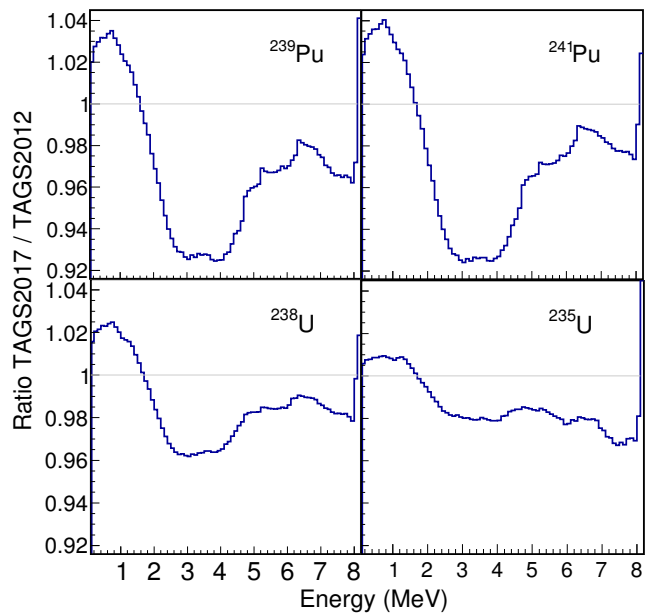}}
\caption{Accumulated impact of the beta intensities of the $^{86,87,88}$Br and $^{91,92,94}$Rb \cite{Valencia,Rice,Zak} decays measured with the total absorption spectrometer {\it Rocinante} on the antineutrino spectra with respect to that published in \cite{Fallot} (relative ratios) for the thermal fissions of $^{235}$U, $^{239}$Pu and $^{241}$Pu, and the fast fission of $ ^{238}$U \cite{Magali_priv}. }
\label{rocinante_cumulated_impact}       
\end{figure}

The cumulative impact of the TAGS beta intensities measured with the {\it Rocinante} detector at Jyv\"askyl\"a on the antineutrino energy spectra generated after the thermal fissions of $^{235}$U, $^{239}$Pu and $^{241}$Pu, and fast fission of $ ^{238}$U are presented in Figure~\ref{rocinante_cumulated_impact} with respect to that built with the most recent evaluation decay databases JEFF3.3~\cite{Jeff} and ENDF/B-VIII.0~\cite{endf} for the same nuclei and containing only the TAGS data from~\cite{Algora,Greenwood}. The decrease of the two plutonium spectra above 1.5~MeV is remarkable, reaching 8\%. The impact on the two uranium isotopes amounts to about 2\% and 3.8\% in the 3 to 4 MeV range in $^{235}$U and $^{238}$U respectively. These results were provided by Dr. M. Estienne \cite{Magali_priv}. 


\begin{figure}
\resizebox{0.48\textwidth}{!}{ \includegraphics{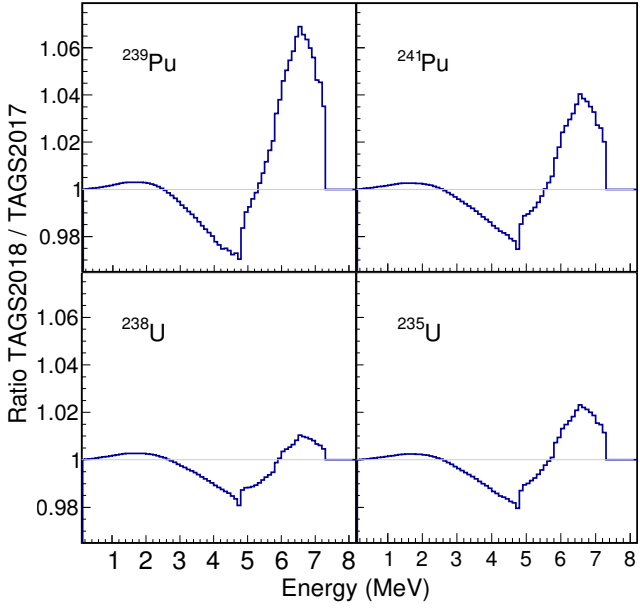}}
\caption{Accumulated impact of the beta intensities measured with the DTAS detector on the antineutrino spectra with respect to that presented in Figure \ref{rocinante_cumulated_impact} (relative ratios) for the thermal fissions of $^{235}$U, $^{239}$Pu and $^{241}$Pu, and the fast fission of $ ^{238}$U \cite{Magali_priv}. The figure represents the relative impact of the $^{100,100m,102,102m}$Nb decays \cite{Guadilla2019}. }
\label{DTAS_cumulated_impact}       
\end{figure}

In our 2014 experimental campaign, we were almost exclusively focussed on nuclei of importance for the prediction of the reactor antineutrino spectrum and for decay heat calculations using the DTAS detector \cite{DTAS}. Twenty-three isotopes were measured, among them many isomers which require the separation power of the Jyv\"askyl\"a Penning trap. An illustration of the experimental challenge is given by the case of the Niobium isomers $^{100, 100m, 102, 102m}$Nb. Nb is a refractory element and the isomers in $^{100}$Nb and $^{102}$Nb are separated  
 by only 313~keV and 94~keV respectively. The half-lives are very similar 1.5~s and 2.99~s in $^{100}$Nb and 4.3~s and 1.3~s in $^{102}$Nb for the ground and isomeric states respectively. $^{100}$Nb and $^{102}$Nb have been assigned a top priority in the list of~\cite{Report-NDS-676}. $^{100}$Nb is among the main contributors to the antineutrino flux in the region of the shape distortion, along with $^{92}$Rb, $^{96}$Y and $^{142}$Cs. 
The results showed that the high resolution measurements for $^{100, 100m}$Nb and $^{102gs}$Nb were affected by the Pandemonium effect, while the beta-intensity distribution for $^{102m}$Nb was determined for the first time~\cite{Guadilla2019}.
The impact of these measurements on the summation calculations was evaluated (see Figure ~\ref{DTAS_cumulated_impact}) and resulted in a large impact between 3 and 7~MeV, with a strong decrease of the spectrum peaked at 4.5~MeV and a strong increase peaked at 6.5~MeV, in the region of the shape distortion. In the calculation, the TAGS data replaced high resolution spectroscopy data extracted from ~JEFF3.3 ~\cite{Jeff} and ENDF/B-VIII.0 ~\cite{endf}. As a result, the discrepancy between the summation antineutrino spectra including these data and the experimental reactor antineutrino spectra is diminished in the region of the shape distortion, though the distortion has not vanished completely ~\cite{Guadilla2019}. The results presented in this last figure were also provided by Dr. M. Estienne \cite{Magali_priv}.


In parallel to the TAGS campaigns, the reactor antineutrino experiments have published their near detector measurement of the emitted antineutrino flux and spectrum from PWRs. In 2017, the Daya Bay experiment could measure the reactor antineutrino flux associated with various fuel compositions~\cite{DayaBay2017}, and found a flux coming from $^{239}$Pu fission in agreement with the prediction of the Huber-Mueller model, while the flux associated with $^{235}$U fission exhibited a deficit of 7\% thus nearly explaining by itself the reactor anomaly. This new result does not favour the idea of oscillation into sterile neutrinos, as it would affect equally the antineutrinos arising from both fuels. It would rather confirm the hypothesis of an additional systematic uncertainty associated with the $^{235}$U energy spectrum. These recent findings reinforced the necessity of an alternative approach to the converted spectra which could be brought by the use of nuclear data. 
It was thus timely to perform a comparison of the summation method spectra with the Daya Bay results. The first comparison was performed in~\cite{Hayes2018} showing a discrepancy with the measured antineutrino flux of only 3.5\%, nearly twice as small as that with the Huber-Mueller model. We have performed an update of our summation model in~\cite{Estienne2019} using the above-mentioned Pandemonium-free datasets improved by the TAGS campaigns of the last decade, the most recent evaluated databases (JEFF3.3, ENDF/B-VIII.0) and updated gross theory spectra~\cite{GrossTheo2018} for the unknown beta decay properties. 
\begin{figure}
\resizebox{0.48\textwidth}{!}{ \includegraphics{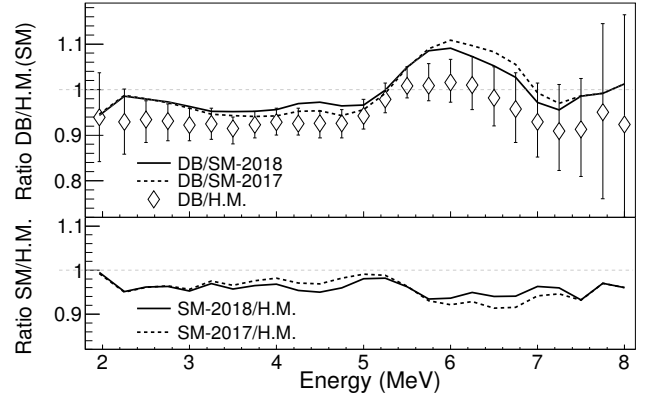}}
\caption{Comparison of the summation antineutrino spectrum obtained using the fission fractions published in~\cite{DayaBay2016} and all the TAGS data quoted in this section, with the experimental spectrum from reference ~\cite{DayaBay2016}. Ratios to the Huber-Mueller (H.M.) model are also provided for comparison. SM-year stands for summation model using the TAGS data analyzed until the given year (see also Figure \ref{Global_impact_Flux_DB} for additional details). Reprinted figure with permission from \cite{Estienne2019}, Copyright (2019) by the American Physical Society.}
\label{Global_impact_Spectrum_DB}       
\end{figure}  

\begin{figure}
\resizebox{0.48\textwidth}{!}{ \includegraphics{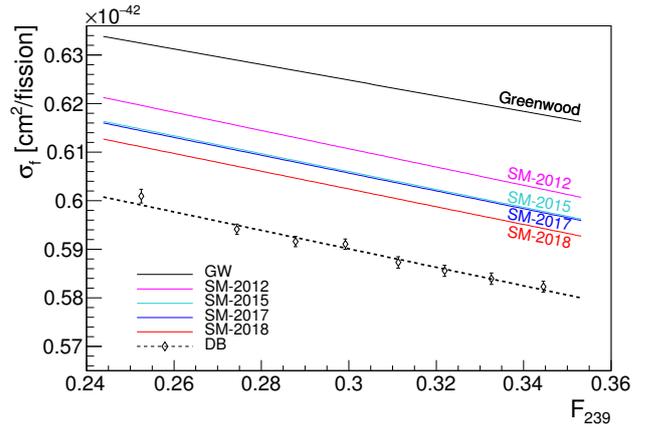}}
\caption{Comparison of the Inverse Beta Decay (IBD) yield computed with the summation antineutrino spectrum obtained using the fission fractions published in~\cite{DayaBay2017} for $^{239}$Pu and using all the TAGS data quoted in this section (included successively), with the experimental IBD yield from~\cite{DayaBay2017}. 
Greenwood represents the result of the summation model \cite{Estienne2019} when only the TAGS results of Greenwood {\it et al.} \cite{Greenwood} are included (for more details of the model see \cite{Estienne2019}). SM-2012 represents the additional impact of the TAGS measurements published in \cite{Algora}  ($^{102,104,105,106,107}$Tc,  $^{101}$Nb and $^{105}$Mo).  SM-2015 contains in addition the effect of $^{92}$Rb \cite{Zak}. SM-2017 represents the impact of  
$^{86,87,88}$Br and $^{91,94}$Rb decays \cite{Valencia,Rice}, and SM-2018 contains the impact of 
$^{100,100m,102,102}$Nb decays \cite{Guadilla2019} (always considered in addition to the earlier version of the summation model). 
Reprinted figure with permission from \cite{Estienne2019}, Copyright (2019) by the American Physical Society.}
\label{Global_impact_Flux_DB}       
\end{figure}  

After folding with the Inverse Beta Decay (IBD) cross section~\cite{BeacomVogel} the summation spectrum built with the actinides spectra weighted with the fission fractions published by Daya Bay, the resulting detected spectrum was compared with that of Daya Bay \cite{DayaBay2016} and that built using the Huber-Mueller model. In Figure ~\ref{Global_impact_Spectrum_DB}, the top panel shows the ratio of the Daya Bay antineutrino spectrum over that computed with the Huber-Mueller model (open diamonds with error bars) superposed with the ratio of Daya Bay over the summation method spectrum including the TAGS results from our first campaign (dashed line) and over the summation method spectrum including the TAGS results from both campaigns (plain line).
The normalisation of the summation method spectrum is clearly in better agreement with the experimental data than that of the Huber-Mueller spectrum (closser to 1), which is considered nowadays the model reference. The inclusion of the TAGS measurements of the Niobium isomers \cite{Guadilla2019} has further improved the shape agreement especially in the energy region of the shape distortion.
The bottom panel of Fig.~\ref{Global_impact_Spectrum_DB} shows the ratio of the summation method spectrum with that of Huber-Mueller. Here again the latest TAGS measurements have flattened the ratio which shows a rather good shape agreement, though located below one at about 95-96\%.
Still the summation method spectrum does not reproduce the shape distortion seen by the reactor antineutrino experiments at PWRs.
Figure~\ref{Global_impact_Flux_DB} summarizes the detected antineutrino flux (called IBD yield) as a function of the fission fraction of $^{239}$Pu obtained with the summation method spectra depending on the TAGS results included in the calculation. The explicit labels of the lines describe the TAGS results introduced one after the other. It is noticable that the inclusion of more TAGS data systematically decrease the detected antineutrino flux to end with an 1.9\% discrepancy with the Daya Bay measured IBD yield. This is a consequence of the correction of the Pandemonium effect in nuclear databases and emphasizes the importance of the TAGS method and measurements.
More details are given in~\cite{Estienne2019} in which the individual IBD yields associated with $^{235}$U, $^{239}$Pu, $^{241}$Pu, and $^{238}$U obtained with the summation model are also compared with the Daya Bay results. The agreement is good in general for all four isotopes. This is at variance with the Huber-Mueller model for which a large discrepancy is observed in the case of $^{235}$U while the three other cases are in very good agreement with the experiment.

In our 2014 TAGS campaign at Jyv\"askyl\"a devoted to reactor antineutrinos and decay heat, $^{96,96m}$Y, $^{140, 142}$Cs, $^{138}$I, $^{137}$I, $^{95}$Rb, $^{95}$Sr, $^{103}$Mo and $^{103}$Tc were measured as well. These future TAGS results may complete the picture that starts to be drawn of the reactor antineutrino energy spectra. In parallel to the nuclear physics effort, reactor antineutrino experiments at short baseline from research reactors start to release their first results. Up to now, only the NEOS and Neutrino-4 collaborations have released  a combined result which signals the presence of an oscillation~\cite{NEOS}. Neither the STEREO~\cite{STEREO} or the PROSPECT~\cite{Prospect} experimental results confirm this oscillation signal. Furthermore, the PROSPECT experiment has released their first spectral measurement of the antineutrino energy spectrum from Highly Enriched Fuel (HEU) which is equivalent to a pure spectrum from $^{235}$U. It is remarkable that their shape-only result does not show such a pronounced shape distortion as the large experiments at PWRs. It thus
excludes the idea that the shape anomaly arises solely from $^{235}$U. It is worth mentioning also that PROSPECT is the first detector using $^6$Li instead of Gd to capture the neutron formed in the IBD process since the Bugey experiment~\cite{Bugey}, which did not see a shape anomaly either. 
Lately Double Chooz has also released their fourth measurement of the $\theta_{13}$ mixing angle obtained by cumulating neutron captures on Gd and the H and C contained in the target and gamma catcher volumes \cite{DCIV}. They observe a shape distortion which could be fitted either with a single or a double Gaussian with a slope. One of their conclusions is that the one sigma envelope for today's prediction appears insufficient to accommodate the mismatch between data and model for both rate and shape. A better understanding of the origin of model deviations remains critical and the role of nuclear data is definitely crucial at the time at which experiments are being set up to measure the mass hierarchy of neutrinos. 
In a recent publication \cite{Berryman}, the global reactor antineutrino data set was re-analyzed using three reactor antineutrino flux predictions, the Huber-Mueller model, the summation method of \cite{Estienne2019} and the model of 
\cite{Hayen} which includes a theoretical calculation of the form factors for the first forbidden transitions. Relative to the traditional Huber-Mueller predictions, the two new calculations result in diverging evidence for a sterile neutrino when total IBD rate measurements are considered. The summation calculation of \cite{Estienne2019} decreases the significance from 2.3  to 0.95 $\sigma$, while that of \cite{Hayen} increases the significance to 2.8 $\sigma$. However, the spectral
anomaly is robust with any of the flux models.
The accurate determination of the reactor antineutrino spectra is also mandatory to monitor future reactors with antineutrino detection. Predictions for innovative reactor designs and fuels can only be obtained through the use of nuclear data and the summation method.
The fine structures present in the antineutrino energy spectra induced by the end-points of the individual beta branches from the fission products~\cite{Dwyer,Sonzo2018} could provide a benchmark for nuclear data and an insight of what is going on inside a reactor. But they also degrade the sensitivity of detectors such as JUNO~\cite{Juno} by mimicking a periodic oscillation pattern. These fine structures may be directly observed by the JUNO-TAO one-ton detector that will be located a few metres away from a PWR core~\cite{Juno-Tao}.  In parallel, an experimental confirmation of the observed first hint of coherent elastic neutrino-nucleus scattering~\cite{COHERENT} would definitely open new possibilities for neutrino applications. 

\section{Nuclear structure}
\label{nuclear}

In an article about the application of TAGS, it would be remiss of us to neglect its application to the study of nuclear structure. It is not our intention here to tell the reader all about beta decay. That can be found in text books (see for example ~\cite{Rubio_euroschool,Krane,Evans,Siegbahn}).
Instead what we want to do is provide a few examples that show how useful TAGS can be in testing nuclear models that contribute to our understanding of the underlying structure of the atomic nucleus, and in particular present some cases recently studied in the framework of reactor applications. 

In Section ~\ref{section_TAS} we already mentioned the $^{100}$Tc case, of relevance for double beta decay studies. Here we will focus on two nuclear structure applications of TAGS that are of significance, namely the study of of the quenching of Gamow-Teller transitions and the study of nuclear shapes.

One essential concept in beta decay, important in the examples that follow,  is the beta strength function (see \cite{Duke}),  a quantity that can be deduced from experiment. It is defined as:  

\begin{equation}\label{exp_strength}
S_\beta(E) = \displaystyle\frac{I_\beta(E)}{f(Q_\beta-E)T_{1/2}}
\end{equation}

where $I_\beta(E)$ is the beta intensity to the level at excitation $E$ in the daughter nucleus, $f$ is the statistical rate Fermi integral which depends on the energy available in the decay ($Q_\beta-E$) and $T_{1/2}$ is the half-life of the decay. $S_\beta(E) $ is in practical terms the reciprocal of the $ft$ values given conventionally in the literature. 
We will concentrate on our contribution to the determination of the strength by measuring the beta feeding in a reliable way, which is the main subject of this article. The other two quantities, namely $Q_\beta-E$ or the $T_{1/2}$ are obtained from measurements dedicated to this purpose. In the following we will focus on allowed Gamow Teller transitions, since allowed Fermi transitions are normally concentrated in a single state, and not affected by $Pande\-monium$ very much. The experimental beta strength is related to the experimental B(GT) in the case of Gamow Teller transitions through the following equation:

\begin{equation}
\label{strength}
S_\beta(E) = \displaystyle\frac{1}{6147} (\frac{g_A}{g_V})^2 \sum_{E} B(GT)^{exp}_{i \rightarrow f}  
\end{equation}


where $g_A$ and $g_V$ are the axial-vector and vector coupling constants. The $B(GT)$ as defined above can be related to the transition probability calculated theoretically between the parent state and the states populated in the daughter defined as follows: 
\\
\begin{equation}
B(GT)_{i \rightarrow f}^{theo}  = | \bra {\Psi_{f}}{\sum_{\substack{
   \mu   
  }} 
  \sum_{\substack{
   k   
  }} 
 \sigma^\mu_{k} \ t^\pm_{k}}\ket {\Psi_{i}} |^{2} 
\label{BGTtheo}
\end{equation}

where $\sigma$ and $t$ represent the spin and isospin operators acting on the individual nucleons and $\Psi_{i}$ and $\Psi_{f}$ the initial and final nuclear states.

In consequence, a comparison of the  $B(GT)^{theo}$ should reproduce the $B(GT)$ determined in the experiment. Actually, the quality of the comparison reflects the goodness of  the nuclear model in describing the involved nuclear states.  
In addition there is a model independent rule, called the Ikeda sum rule, that tells us how much strength we should observe. Curiously enough, the strength obtained experimentally seems to be systematically lower than theory. This is called the Gamow Teller quenching problem and it has been discussed for more than four decades and is not yet fully resolved (see for instance ~\cite{Ichimura,Pinedo} and more recently \cite{Nature_quenching} and references therein). The discussion of this mismatch between theory and experiment involves theoretical as well as experimental arguments. One main difficulty is that the full strength, in the case of GT transitions, is normally concentrated in a resonance at relatively high excitation energy, for example in the range of 8-15 MeV for A$\sim$100, that is difficult to access in beta decay. In consequence, charge-exchange reactions with hadronic probes, such as (p,n), ($^{3}$He,t) or (t,$^{3}$He),  have been used to measure the full strength. Extracting the GT strength from these probes is more complicated than extracting it from beta decay. Among other reasons, this is because it relates to the penetrability of the hadronic probe in the nucleus and because of the difficulty of selecting the GT process in a clean way.
Beta decay, however, has its own difficulties. The principal one, as mentioned above, is to know how much of the strength lies within the $Q$-value window, and the other difficulty is to be sure that we measure all the strength inside the $Q_\beta$ window, because of the Pandemonium effect. It is in overcoming this second difficulty that the TAGS technique has had an impact in tackling this problem.
\\

In order to avoid the first problem, one can choose cases where most of the strength is expected to be located at relatively low energy, inside the $Q_\beta$ window, and this happens, in principle, in the beta decay of nuclei south east of $^{100}$Sn on the Nuclear Chart and in the rare-earth nuclei above the spherical nucleus $^{146}$Gd. These cases, even although they do not have a direct relation with the cases of relevance for reactor applications presented in this article, can provide information on the necessary corrections that are required for a proper theoretical description of the beta decay process. The reason, in both the $^{100}$Sn and above $^{146}$Gd regions, is that there is only one main component in the GT strength on the $\beta^{+}$ side, namely $\pi$g$_{9/2}$$\rightarrow$$\nu$g$_{7/2}$ in the $^{100}$Sn region and $\pi$h$_{11/2}$$\rightarrow$$\nu$h$_{9/2}$ in the rare-earth case. All other proton occupied orbitals have no empty neutron orbital partner. Unfortunately, the expected B(GT) cannot be directly compared with the Ikeda sum rule in these cases because this rule involves the B(GT) values for both the $\beta^{+}$ and the $\beta^{-}$ decays and here only the former can be measured. So, it has to be compared with theory. A relatively simple but realistic calculation of the expected beta strength on the $\beta^{+}$ side was carried out by Towner ~\cite{Towner} for decays in both of these regions of the Segre Chart. In this work ~\cite{Towner} a hindrance factor $h$ is defined as the ratio between the summed GT strength from theory and experiment. Initially he adopted  the extreme single particle approach ($s.p$), namely considering only the two pairs of orbitals, $\pi$g$_{9/2}$-$\nu$g$_{7/2}$ and $\pi$h$_{11/2}$-$\nu$h$_{9/2}$. He then made a series of corrections to this approach taking into account pairing, core polarization and higher-order effects and then looked at how hindered the corrected theoretical strength  would be in comparison with the extreme $s.p$ picture. This result defines a theoretical hindrance factor that can be compared later with the hindrance obtained from the ratio of the extreme single particle approximation and experiment. The theoretical hindrance  
was calculated for the range of cases from n=1 to n=10 active protons in the g$_{9/2}$ orbital in the $^{100}$Sn region and n=1 to n=12 in the h$_{11/2}$ orbital in the rare earths.
\\

\begin{figure}
\resizebox{0.53\textwidth}{!}{ \includegraphics{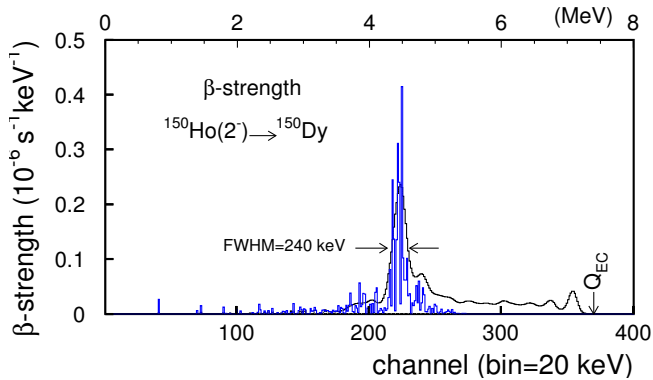}}
\caption{Comparison of the beta strength deduced from the high resolution measurement of the beta decay of the $^{150}$Ho $2^{-}$ isomer using the cluster cube setup \cite{Algora_clus}  (shown in blue) 
with the strength obtained from a total absorption measurement \cite{Cano_150} (black) . The measurements were performed at the Mass Separator at GSI. Reprinted figure with permission from \cite{Algora_clus}, Copyright (2003) by the American Physical Society.}
\label{cluster_cube}       
\end{figure}

 A series of experiments were carried out at GSI (Germany) with heavy ion beams from the UNILAC at energies slightly above the Coulomb barrier on the appropriate targets to study relevant beta decays in these regions of interest. At these energies, the reaction was dominated by the fusion evaporation channels which are fewer than in the fission examples described in previous sections. Consequently, the separation achieved with the relatively simple  Mass SEParator (MSEP) ~\cite{MSEP}, provides clean enough samples to perform the experiments. The GSI TAS ~\cite{KarnyTAS} was built and briefly used at the Berkeley SuperHILAC and after the accelerator was closed,  it was installed at the GSI MSEP.  This spectrometer enjoyed two advantages over the INEL Idaho TAGS ~\cite{Greenwood_MC}. Firstly, the crystal was more than twice the size, secondly it included a small cooled high purity Ge detector for X-Ray detection. The first improvement was important for a better response of the spectrometer to the absorption of the gamma cascades, and the second to clean the EC (Electron Capture) component of the decays further and hence to obtain a very good Z separation. The results can be seen in references  ~\cite{Plettner} ($^{100}$In), ~\cite{Hu} ($^{97}$Ag), ~\cite{Algora_Dy} ($^{148}$Dy) ~\cite{Algora_clus,Cano_150} ($^{150}$Ho) and 
~\cite{Nacher2} ($^{148}$Tb(2$^{-}$ and 9$^{+}$ isomers), and $^{152}$Tm(2$^{-}$ and 9$^{+}$ isomers)). For the very special case of $^{100}$Sn, where the expectations are that all the strength is concentrated in a a single 1$^{+}$ state in the daughter, we will consider  the results of Hinke {\it et al.} ~\cite{Hinke}  and the more recent work by Lubos {\it et al.} \cite{Lubos}
measured with Ge detectors. 
To study this case further, a TAGS experiment has been proposed and partially carried out at RIKEN ~\cite{RIKEN_Proposal} and is currently under analysis \cite{Algora-annualreport2}. In general, one observes that the calculations ~\cite{Towner} reproduce the observed hindrance factor fairly well for the $^{100}$Sn region.  
One should mention here that this case is simple enough that it has been calculated from first principles reproducing the experimental value also fairly well ~\cite{Nature_quenching}. 
At the time of this last work \cite{Nature_quenching} only the ~\cite{Hinke} results were available and included in the comparison, but the more recent results of \cite{Lubos} show even better agreement with these calculations, similar in quality to the agreement with the systematic extrapolation of the strength from \cite{Batist}. 
In contrast, in the rare-earth region higher hindrance factors have been observed experimentally compared to Towner's calculations. One should note that, somewhat surprisingly, the GT resonance is clearly observed in this case. Moreover, the tail below the resonance that has been observed experimentally and discussed at length in the Charge Exchange reaction experiments, is clearly seen in these beta decay experiments for the first time. 
As an example, we show in Figure \ref{cluster_cube} the case of the decay of the $^{150}$Ho 2$^{-}$ isomer to $^{150}$Dy \cite{Algora_clus}, with the TAGS measurement represented by the spectrum under the black line. A similar result was obtained in the $^{152}$Tm case reported in ~\cite{Nacher2}. In this article, it was suggested that the missing strength, or in other words the explanation of the disagreement with Towner, is probably located in that part of the tail of the resonance which is cut off by the  $Q_\beta$ window. 
\\
\\
The same Figure~\ref{cluster_cube} can also be used to illustrate the importance of the TAGS experiments and the limitations of the Ge detectors in terms of observing beta feeding at high excitation energy that was discussed in the introduction.  We see the B(GT) distribution measured with the Cluster cube ~\cite{Algora_clus} and with the GSI TAS ~\cite{Cano_150}. The Cluster cube was an array of six Euroball Ge cluster detectors in compact geometry. It was equivalent to forty two individual Ge detectors and had an efficiency of 10.2(5) $\%$ at a gamma-ray energy of 1332 keV. The figure shows clearly the importance of both types of measurement. In the Ge measurements 1064 gamma rays were identified and the coincidences between detectors allowed the construction of a decay scheme with 295 levels in $^{150}$Dy \cite{Algora_clus}. Figure~\ref{cluster_cube} shows in blue  the B(GT)  strength to each of these levels as deduced from the beta feedings in the decay scheme. Inspection of the total absorption  spectrum reveals that the Ge array loses sensitivity as a function of excitation energy in the daughter nucleus when compared with the TAGS and our ability to determine the feeding or beta intensity distribution diminishes. Once converted into B(GT) strength we conclude that we lose 50$\%$ of the total strength compared with the TAGS measurement. Moreover, the tail of the resonance, one of the foci of interest in the experiments discussed above cannot be seen with the Ge detector array. 
\\

In summary, returning to the discussion of the missing B(GT) strength in beta decay, even though the experiments explained above are probably the best cases to study the Gamow Teller quenching, they rely very much on comparison with theoretical calculations which, in general, cannot locate with sufficient precision whether the calculated strength is inside the accessible beta window or not. However, these measurements have demonstrated that the tail below the Gamow Teller resonance exists, and this observation is free of background ambiguities. Moreover, in the future, when either the calculations are accurate enough to tell us how much of the strength should be located within the accessible beta-window, or, alternatively when we are able to perform charge exchange reactions using radioactive beams, we have here very reliable measurements of that part of the spectrum that lies below the $Q_\beta$ energy.  This can be used for normalisation purposes as well as for control of the reaction mechanism. 
\\

Another application of TAGS measurements, first pioneered at CERN-ISOLDE with the Lucrecia TAGS, relates to the shapes of nuclear ground states. The concept of nuclear shape is deceptively simple. In practice it is difficult to measure. The measurements with TAGS are based on a theoretical idea put forward by Hamamoto and Zhang ~\cite{Hamamoto}, that was developed further by Sarriguren {\it et al.} ~\cite{Sarriguren} and  Petrovici  {\it et al.} ~\cite{Petrovici}. They showed that the beta strength distribution for transitions to excited states in the daughter nucleus depends on the shape assumed for the ground state of the decaying nucleus. Intuitively one can see why this might be so if we look at the ordering of deformed single particle orbits on a Nilsson diagram. The levels on the prolate and oblate sides are in different order and thus filling them up to the Fermi level to determine the ground state configuration of a particular nucleus involves different single particle contributions. Their beta decay strength distributions will also be different since they are dictated by angular momentum and parity selection rules as well as the overlap of the wavefunctions of the states involved. The calculations by Hamamoto and Zhang, Sarriguren {\it et al.} and Petrovici {\it et al.}, are of course rather more sophisticated than this simple picture which is just used for understanding the underlying physics phenomena.  

In particular regions of the nuclide chart, nuclei can have several minima in the potential energy surface with different shapes for the ground state. The calculated B(GT) distributions for each of these states with a defined deformation are quite different in some but not all cases (see for example ~\cite{Sarriguren}). Where they are different,  the experimental B(GT) distribution measured with TAGS can then be compared with the theoretical distributions and the ground state shape inferred. A number of studies of this kind (\cite{Nacher}, \cite{Poirier}, \cite{Perez},  \cite{Briz}, \cite{Aguado}),  have been carried out for nuclei with A$\sim$ 80 and A$\sim$190. A summary of these activities at ISOLDE can be found in ~\cite{Rubio}.  

\begin{figure}
\resizebox{0.48\textwidth}{!}{ \includegraphics{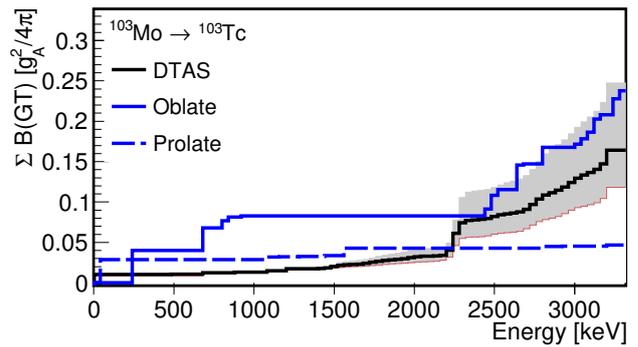}}
\caption{Comparison of the deduced beta strength for the decay of $^{103}$Mo  \cite{Guadilla-thesis,Guadilla_inprep} in comparison with QRPA calculations assuming prolate or oblate deformations in the ground state of $^{103}$Mo \cite{Sarriguren_priv}. In the model a quenching factor of $(\frac{g_A}{g_V})_{eff}=0.77(\frac{g_A}{g_V})$ is applied.}
\label{103Mo}       
\end{figure}

This method has also been applied for some of the cases studied at IGISOL. See for example $^{100,102}$Zr  and $^{103}$Mo \cite{Guadilla-thesis,Guadilla_PRCNb}. In Figure \ref{103Mo} \cite{Guadilla-thesis,Guadilla_inprep} we present a comparison of the deduced strength for the decay of $^{103}$Mo with Quasiparticle Random Phase Approximation (QRPA) calculations performed by P. Sarriguren \cite{Sarriguren_priv}. From this comparison a preference for an oblate shape in the ground state of $^{103}$Mo can be inferred. In the calculations a quenching factor of 
$(\frac{g_A}{g_V})_{eff}=0.77(\frac{g_A}{g_V})$ has been applied, which is equivalent to a hindrance factor of 1.69 with respect to the QRPA calculations used in the comparison.


\begin{figure}
\resizebox{0.48\textwidth}{!}{ \includegraphics{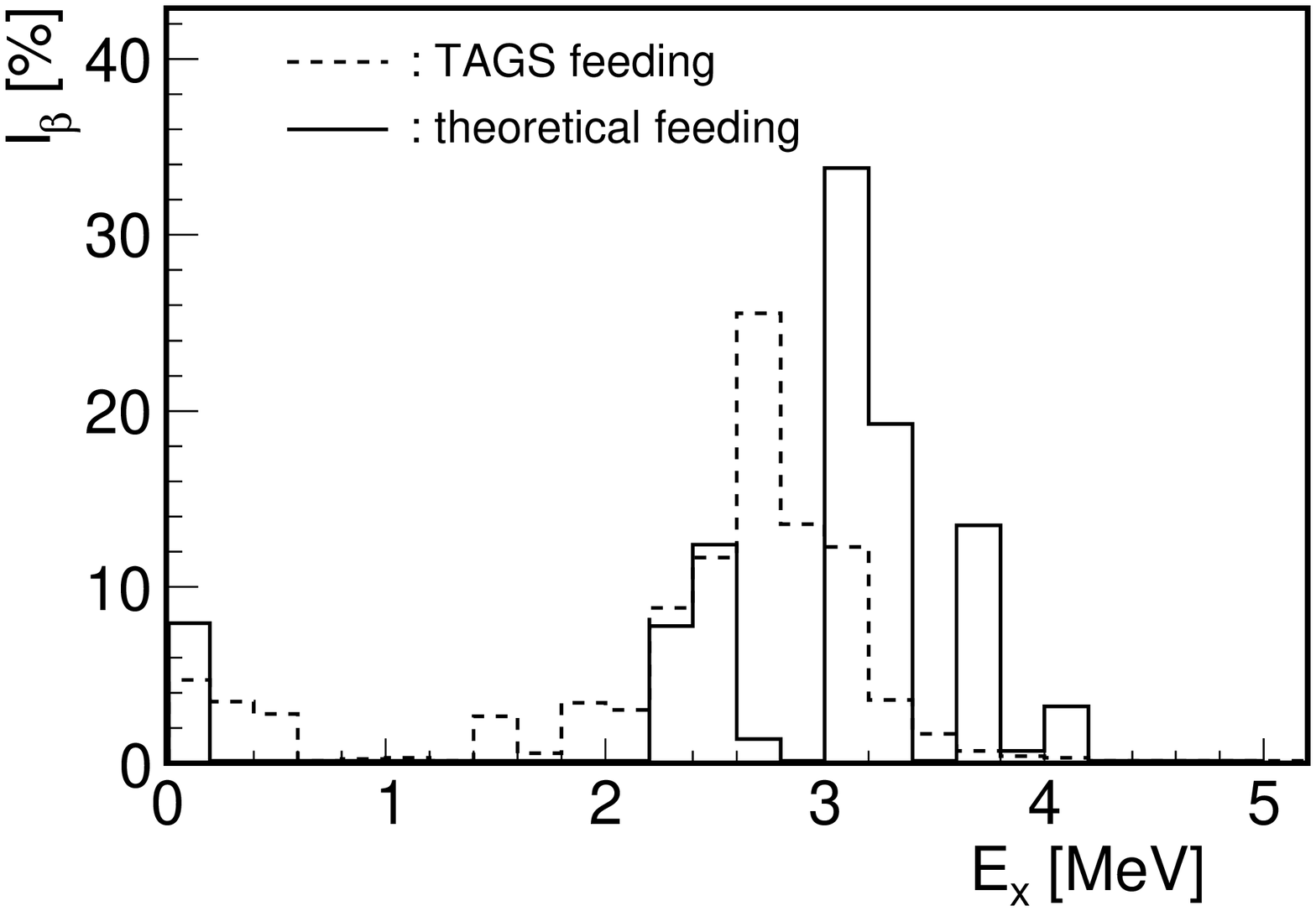}}
\vskip -.6 cm
\resizebox{0.48\textwidth}{!}{ \includegraphics{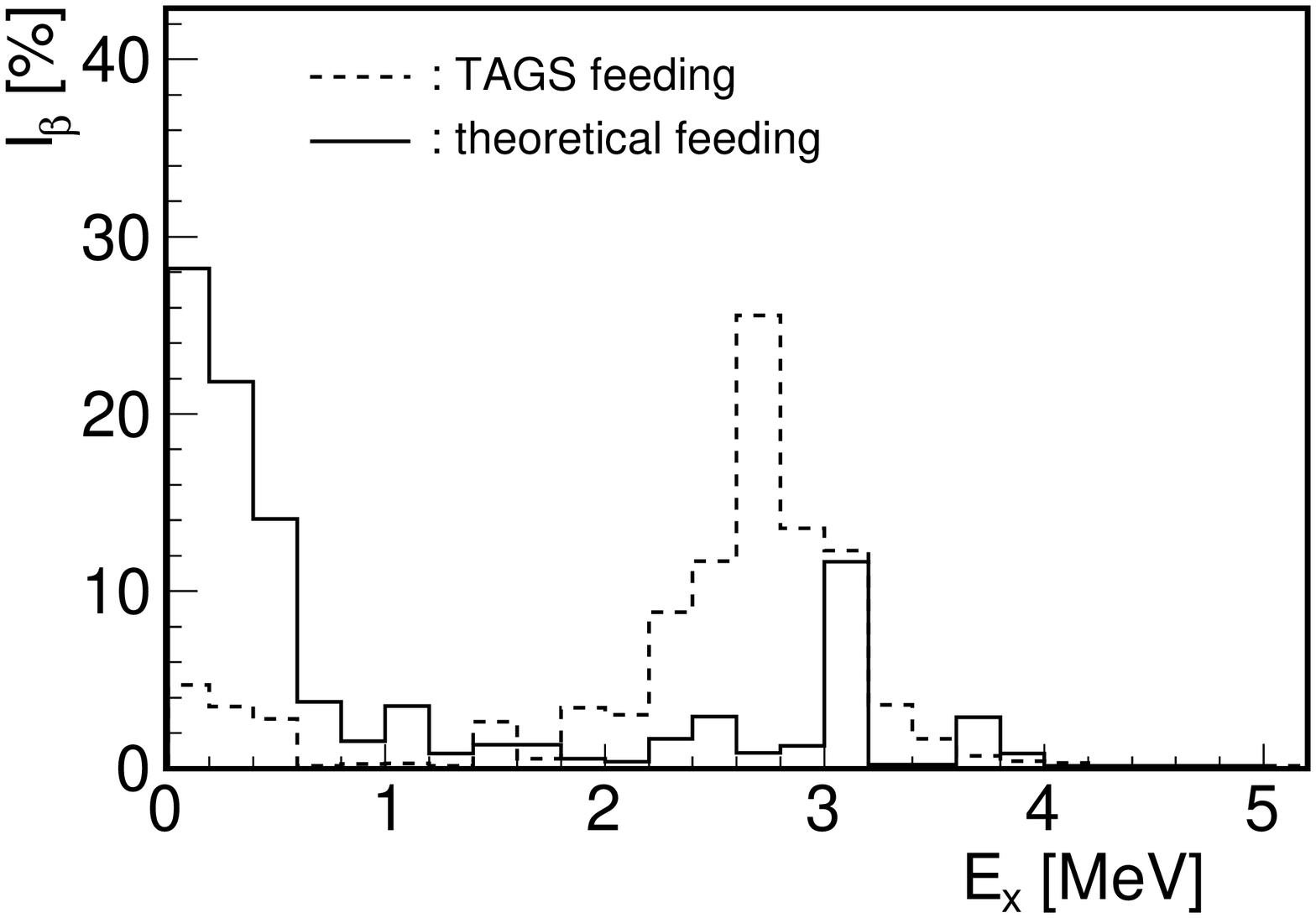}}
\caption{Comparison of the experimentally deduced beta feeding in the decay of  $^{105}$Mo with the results of theoretical calculations using the FRDM-QRPA model \cite{Kratz-Moller,Jordan-2}. The upper 
panel is obtained assuming only allowed GT transitions. The lower panel shows the comparison with calculations that also include the first forbidden component. See more details in the text.}
\label{105Mo}       
\end{figure}

Another example of the importance of TAGS measurements in testing models is provided  by $^{105}$Mo. The decay of $^{105}$Mo was calculated using the FRDM-QRPA model \cite{Kratz-Moller,Jordan-2}. The best theoretical description of this decay was obtained assuming a ground state deformation of $\epsilon_{2}$=-0.31 for $^{105}$Mo. The experimental half-life of this decay is 35.6 s, and this value can be better reproduced if first forbidden transitions are included in the model calculation ($T_{1/2}^{theo}$=30.3 s), but in that case, the experimental beta distribution is not  reproduced so well. This can be seen in Figure \ref{105Mo} where the experimental feeding distribution is compared with the theoretically deduced distributions with and without first forbidden transitions. A better reproduction of the beta distribution by theory is obtained if no first forbidden component is included in the model. But in that case the experimental half-life is not so nicely reproduced ($T_{1/2}^{theo}$=150 s). This clearly shows a limitation of the performance of this model in a region which is dominated by shape effects and where triaxiality can play a role. QRPA  calculations assume that both the parent and the daughter have the same deformation, which might not always be applicable in regions where shape transitions are common. This example shows the relevance of having in addition to the experimental half-life the possibility of comparing the theoretical strength (or the deduced theoretical feeding) with reliable experimental data, like that provided by TAGS measurements. Based on the description of the half-life only, we might conclude it is necesary to introduce the first forbidden component for the description of this decay, which does not reproduce well the experimental beta feeding. The relevance of this kind of model validation will be further discussed in the next Section in relation to astrophysical applications. 


\section{Astrophysical applications}
\label{astrophysics}

TAGS measurements are important also in the context of nuclear astrophysics. We select here some examples related to the astrophysical {\it r} process. As mentioned in the previous Section, some of those examples will show a strong interrelation with nuclear structure studies. 

The {\it r} process is driven by a huge instantaneous flux of neutrons that creates by successive neutron captures very neutron-rich nuclei, up to the heaviest ones, that then beta decay towards stability. About half of the observed abundance of elements heavier than Fe in the Universe is synthesized in this way. The identification of the astrophysical site where the process occurs is the subject of very active investigations. Core Collapse Supernovae were the classical favoured site in spite of persistent difficulties met when trying to reproduce observations with calculations. On the other hand Neutron Star Mergers have recently become \cite{Watson} a confirmed site for heavy element formation after the first observation of the gravitational waves generated and the analysis of the subsequent electromagnetic radiation. Much remains still to be done in order to understand the role of both scenarios combining astrophysical observations and calculations that require nuclear physics input (see \cite{Horowitz} for a recent review).

Some of the key input parameters in {\it r} process calculations are the decay properties of  neutron-rich nuclei, more specifically half lives ($T_{1/2}$) and beta-delayed neutron emission probabilities ($P_{n}$) that control the nucleosynthesis flow. For such exotic nuclei the neutron separation energy ($S_{n}$) becomes smaller than the decay $Q_\beta$ value and neutron emission from populated neutron unbound states occurs. In spite of current efforts at the most advanced radioactive beam facilities to determine this information experimentally \cite{Tain_Mazurian}, most of the nuclei involved cannot be accessed in the laboratory and need theoretical estimates. The key point here is that both quantities are derived from the beta strength distribution $S_{\beta} (E)$ (see also Equation~\ref{exp_strength}) 

\begin{eqnarray}
\frac{1}{T_{1/2}} & = & \int_{0}^{Q_{\beta}} S_{\beta} (E) f(Q_{\beta}-E) dE 
\label{halflife} \\
P_{n} & = & T_{1/2} \int_{S_{n}}^{Q_{\beta}} \frac{\Gamma_{n}}{\Gamma_{n}+\Gamma_{\gamma}} S_{\beta} (E) f(Q_{\beta}-E) dE 
\label{neutprob}
\end{eqnarray}

Equation \ref{neutprob} above includes the competition between neutron ($\Gamma_{n}$) and gamma ($\Gamma_{\gamma}$) emission. It should be noted that models often assume that neutron emission prevails always and the competition is ignored.

The quality of beta strength calculations is usually asserted by global comparisons with measured half-lives and to a lesser extent with measured neutron emission probabilities. However it is found that different theoretical models with comparable quality predict quite different $T_{1/2}$ and $P_{n}$ (see for example \cite{Caballero}). This clearly indicates that the quality assessment based on these integral quantities (Equations \ref{halflife}  and \ref{neutprob})  is not good enough. This comes as no surprise since several (theoretical) strength distributions can lead to the same half-life for a particular nucleus. But the underlying nuclear structure can then predict very different numbers for neighboring nuclei. Comparing TAGS measurements of the beta strength with different models then becomes the only reliable validation method (see the $^{105}$Mo case discussed in the previous Section). 
Moreover it can give us hints on how to improve the nuclear structure calculations. 

Another example is related to the determination of neutron capture $(\mathrm{n},\gamma)$ cross-sections for very exotic neutron-rich nuclei, that also controls the nucleosynthesis flow in the  {\it r} process. These are even more difficult to determine experimentally because of the need to prepare suitable targets. Direct measurements will require very imaginative techniques thus current efforts concentrate on indirect methods \cite{Larsen}. 
Theoretical estimates are based on the statistical Hauser-Feshbach model \cite{Hauser} that uses average quantities: nuclear level densities, photon strength functions and neutron transmission coefficients. These are parameterized using data measured mostly close to stability and consequently there is considerable uncertainty on the values needed in {\it r} process calculations.

We have proposed a way to obtain experimental constraints on the quantities that intervene in the Hauser-Feshbach estimate for very exotic neutron rich nuclei \cite{Tain,Valencia}. It is based on the analogy between radiative neutron capture reactions and the process of beta-delayed neutron emission. The former depends mostly on $\Gamma_{\gamma}$ and weakly on $\Gamma_{n}$ and the latter can provide the ratio $\Gamma_{\gamma}/\Gamma_{n}$ provided that we are able to measure the (expected weak) gamma emission from neutron unbound states. This is where the sensitivity of the TAGS technique comes into play. The main advantage of the method is that the measurements can be extended into regions quite far from stability.



In \cite{Tain,Valencia} the gamma-neutron competition was studied for the $^{87,88}$Br and $^{94}$Rb decays and more recently in \cite{Guadilla_neutron} for 
$^{95}$Rb and $^{137}$I . The results are summarized in Table \ref{Pgamma}, which shows $P_\gamma$,  the gamma emission probability above $S_{n}$ defined by analogy with $P_{n}$ (also shown). Observation of Table \ref{Pgamma} reveals that in most of the cases $P_\gamma$ is large, even larger than $P_{n}$. The large $P_\gamma$ for $^{137}$I is confirmed by the TAGS measurement of \cite{Rasco}. The reason for this surprising result is to be found in the nuclear structure of the nuclei in the decay chain. A large mismatch between spin and parity of unbound states in the daughter nucleus and the available states in the final nucleus means that neutron emission is hindered by the centrifugal barrier. Other measurements have also found large $P_{\gamma}$ values in the decay of $^{70}$Co \cite{Spyrou} and $^{83}$Ga \cite{Gottardo} and different nuclear structure effects were invoked to explain it. This notable result warns us about the neglect of gamma-neutron competition in theoretical estimates of $P_{n}$ (see also \cite{Mumpower}), but does not tell us about the statistical parameters of the Hauser-Feshbach model. The most interesting case from that point of view is the decay of $^{94}$Rb where level densities in the daughter nucleus are large and neutron emission is not hindered by angular momentum mismatch, making it a good test case for the statistical model. In this case gamma emission above $S_{n}$ represents only 5\% of neutron emission but even so it is more than one order-of-magnitude larger than Hauser-Feshbach calculations using standard statistical parameters. This is a challenging outcome. However in order to translate a constraint in $\Gamma_{\gamma}/\Gamma_{n}$ into a constraint on $(\mathrm{n},\gamma)$ cross-section we need additional information. If we follow the assumption that extrapolating far from stability nucleon optical parameters (that determine $\Gamma_{n}$) is more reliable than extrapolating photon strength functions (that determine $\Gamma_{\gamma}$) then this result would indicate one order-of-magnitude increase in the calculated capture cross-section. Clearly more investigations are required and new TAGS measurements on suitable isotopes are planned.

A different method to obtain constraints on $(\mathrm{n},\gamma)$ cross-sections for unstable nuclei using TAGS measurements has been proposed by the NSCL group \cite{Spyrou_Oslo}. It was already mentioned in Section \ref{section_TAS} in connection with the extraction of the branching ratio matrix, which is the first step of the Oslo method \cite{Oslo}. The goal of the Oslo method is to obtain the shape of the nuclear level density and photon strength function from the branching ratio matrix (in their terminology: primary gamma ray intensities) and it was originally applied to nuclear reaction experiments. Going from a relative quantity (branching ratios) to absolute quantities (photon strength functions and nuclear level densities) requires the use of normalization parameters coming from external sources. In the case of beta decay TAGS measurements away from the stability these are systematics, extrapolations or theory. Thus the impact of this method is mainly that of the shape of the photon strength function.


A closely related topic and also very interconnected with nuclear structure is the potential provided by beta decay in relation to the study of collective phenomena. 
Beta decay could constitute a new means to investigate the presence and maybe some of the properties of low-lying collective modes, such as pygmy dipole modes predicted to appear at lower energies as nuclei become more neutron rich. Collective modes are of crucial importance in nuclear structure as they reflect the ability of the nucleons to move  
coherently and provide insights into the properties of the nuclear force. The study of collective modes puts constraints on theoretical models as well. They are also the only observables that we can study on earth providing access to the intrinsic properties of nuclear matter, entering into the modelling of astrophysics phenomena like supernovae or neutron stars. Pygmy dipole resonances (PDR) could be the consequence of the appearance of neutron skins in medium to heavy neutron-rich nuclei. The  PDR might deliver information on neutron-star properties 
\cite{Horowitz2}. Important information on the equation of state (EOS) of neutron-rich matter via strength-neutron-skin thickness correlation could be obtained \cite{Piekarewicz}. 

The presence of low-lying PDR could influence processes of nucleosynthesis, especially (n,$\gamma$), ($\gamma$,n) reactions playing an important role in the r-process \cite{Goriely} as mentioned earlier. Several questions remain  unanswered about the collective modes when nuclei become more exotic. One limitation up to now has been the low intensity of the accessible exotic beams which limits the possible studies using standard nuclear or electromagnetic probes.  In this context beta decay constitutes a new probe for low-lying collective modes. Further away from stability, as the energy window opened by beta decay increases, the energy of the pygmy modes decreases, allowing their excitation through the Gamow-Teller operator when the spin and parity conservation conditions are fulfilled. Beta decay then offers new possibilities to study systematically the presence of low-lying collective modes with the existing exotic beam intensities. Our collaboration was first to propose an experiment on this topic \cite{Alto_prop}. Later on, the theoretical demonstration was provided by two models \cite{Gottardo,Scheck}. The quasi-particle model of \cite{Scheck}  predicts that other components of the collective mode are excited through beta decay than those excited by the usual nuclear and electromagnetic probes. In particular, beta decay would feed preferentially two-particle two-hole components of the collective mode, being thus complementary to nuclear reactions. In the experimental results of \cite{Scheck} and \cite{Gottardo}, 
high resolution setups with a relatively small detection efficiency were used and the data may suffer from the Pandemonium effect. The TAGS technique, using modern segmented spectrometers, seems to be very well adapted to tackle this problem, especially to evidence high energy gamma-rays feeding the daughter ground state or the first excited state. In parallel,  ways to obtain experimental evidence of the collectivity of the states fed by beta decay, which would not rely on theoretical predictions, should also be investigated.

\begin{table}
\caption{P$_\gamma$ obtained from our measurements \cite{Valencia,Guadilla_neutron} in comparison with the Pn values of the decays. P$_\gamma$ is defined as the gamma emission probability above the $S_n$ value (in analogy to $P_n$). The values are given in \% (see the text for more details).  }
\label{Pgamma}       
\begin{center}
\begin{tabular}{c  c c }
\hline\noalign{\smallskip}
  Isotope         & $P_\gamma (TAGS)$ & $P_n $\\
\noalign{\smallskip}\hline\noalign{\smallskip}
   $^{87}$Br	                                  &   3.50$^{+0.49}_{-0.40}$        &  2.60(4)      \\
   $^{88}$Br	                                 &   1.59$^{+0.27}_{-0.22}$       &   6.4(6)       \\ 
   $^{94}$Rb	                                 &   0.53$^{+0.33}_{-0.22}$       &   10.18(24) \\ 
   $^{95}$Rb	                                &   2.92$^{+0.97}_{-0.83}$       &  8.7(3)        \\    
   $^{137}$I	                               &   9.25$^{+1.84}_{-2.23} $        &  7.14(23)     \\  
\noalign{\smallskip}\hline
\end{tabular}
\end{center}
\end{table}







\section{Summary, future and conclusions}
\label{summary_future}

In this article we have presented a review of the impact of our total absorption studies of beta decays that are relevant for reactor applications. The measurements presented have been performed at the IGISOL facility of the University of Jyv\"askyl\"a employing the high isotopic purity beams provided by the JYFL Penning Trap. These measurements are not only relevant for the decay heat predictions and for the predictions of the reactor neutrino from reactors, but also provide results of interest for nuclear structure and astrophysics. In particular they offer the possibility of testing nuclear models in a more stringent way and can provide additional information for the estimation of (n,$\gamma$) cross sections of astrophysical interest for cases not directly accessible using reactions. 

Considerable progress has been made, but the ultimate goal of the work presented in this article has not yet been reached. From the comparisons of the measured decay heat with the predictions of summation calculations, it is clear that there is still work to be done, in particular for the $^{235}$U fuel. The situation is similar in relation to the prediction of the antineutrino spectrum in reactors, where the remaining discrepancies still require to measurements of a number of decays. Our collaboration is still working on these subjects and has approved proposals to continue our studies at the IGISOL IV facility in order to measure new decays that are important in the next relevant order. In this publication we have concentrated mainly on the discussion of results obtained by our collaboration, but other groups are also involved in similar research programmes at other facilities that provide experimental results relevant to the topics discussed here (see for example \cite{Rasco,Rasco_let,Alexa,Spyrou}). The upgrade and advent of a new generation of radioactive beam facilities like FRIB (Michigan, USA), RIBF (RIKEN, Japan), FAIR (Germany), Spiral2 (France), etc. extends the possibilities of TAGS measurements to more exotic domains than those offered by the present and future ISOL facilities. These measurements represent new challenges concerning the purity of the beams and require the development of detectors adapted to the experimental conditions of such a facilities. From those facilities new and exciting results will appear in the near future.

This work has been supported by the Spanish Ministerio de Econom\'ia y Competitividad under Grants No. FPA2011-24553, No. AIC-A-2011-0696, No. FPA2014-52823-C2-1-P, No. FPA2015-65035-P, FPA2017-83946-C2-1-P, No. FPI/BES-2014-068222 and the program Severo Ochoa (SEV-2014-0398), by the Spanish Ministerio de Educaci\'on under the FPU12/01527 Grant, by the European Commission under the European Return Grant, MERG-CT-2004-506849, the FP7/EU\-RATOM contract 605203 and the FP7/ENSAR contract 262010, and by the $Junta~para~la~Ampliaci\acute{o}n~de~Estudios$ Programme (CSIC JAE-Doc contract) co-financed by FSE. We acknowledge the support of STFC(UK) council grant ST/P005314/1. This work was supported by the CNRS challenge NEEDS and the associated NACRE project, as well as the CHANDA FP7/EURATOM project (Contract No. 605203),  and SANDA project ref. 847552, the CNRS/\-in2p3 PICS TAGS between Subatech and IFIC, and the CNRS/in2p3 Master projects Jyv\"askyl\"a and OPALE. 
Thanks are also due to all collaborators who participated in the measurements, to the IGISOL and University of  Jyv\"askyl\"a colleagues for their continous support and help and in particular to the PhD students and colleagues who worked in the analysis of the data and made this work possible (D. Jordan, E. Valencia, S. Rice, V. M. Bui, A. A. Zakari-Issoufou, V. Guadilla, L. Le Meur, J. Briz-Monago and A. Porta). We also thank A. L. Nichols and T. Yoshida for their support in the earlier stages of the work and P. Sarriguren, A. Petrovici, K. L. Kratz, P. M\"oller and collaborators for providing theoretical calculations for some of the cases studied.
Thanks are also due to A. Sonzogni and L. Giot  for providing decay heat calculations and to M. Estienne for providing antineutrino summation calculations. The work of J. Agramunt in the development of our data acquisition system used in all the experiments is acknowledged. Support from the IAEA Nuclear Data Section is acknowledged. 

%

%
%
\section{Authors contributions}
All the authors were involved in the preparation of the manuscript.
All the authors have read and approved the final manuscript.
%

\begin{thebibliography}{}
%
%
\bibitem{Pauli} W. Pauli letter to nuclear physicists in Tuebingen, Germany, 1930.
\bibitem{Fermi} E. Fermi, Il Nuovo Cimento, Volume 9, 1 (1934).
\bibitem{Hardy} J. C. Hardy {\it et al.}, Phys. Lett. 71B, 307 (1977).
\bibitem{Hu_98} Z. Hu {\it et al.}, Phys. Rev C 62, 064315 (2000).
\bibitem{Nacher} E. Nacher {\it et al.}, Phys. Rev. Lett. 92, 232501 (2004).
\bibitem{Poirier} E. Poirier {\it et al.}, Phys. Rev. C 69, 034307 (2004). 
\bibitem{Algora} A. Algora {\it et al.}, Phys. Rev. Lett. 105, 202501 (2010).
\bibitem{Jordan} D. Jordan {\it et al.}, Phys. Rev. C 87, 044318 (2013).
\bibitem{Perez} A. B. Perez-Cerdan {\it et al.}, Phys. Rev. C 88, 014324 (2013). 
\bibitem{Briz} J. A. Briz {\it et al.}, Phys. Rev. C 92, 054326 (2015).
\bibitem{Aguado} M. E. Estevez Aguado {\it et al.}, Phys. Rev. C 92, 044321 (2015).
\bibitem{Tain} J. L. Ta\'in  {\it et al.}, Phys. Rev. Lett. 115, 062502 (2015).
\bibitem{Guadilla2019} V. Guadilla  {\it et al.}, Phys. Rev. Lett. 122, 042502 (2019).
\bibitem{Duke} C. L. Duke {\it et al.}, Nucl. Phys. A151, 609 (1970).
\bibitem{Rubio2005} B. Rubio {\it et al.}, Journal of  Phys. G: Nucl. Part. Phys. 31, S1477 (2005).
\bibitem{Rubio} B. Rubio {\it et al.}, Journal of  Phys. G: Nucl. Part. Phys. 44, 084004 (2017). 
\bibitem{Greenwood} R. C. Greenwood {\it et al.}, Nuclear Instruments and Methods in Physics Research A 390, 95 (1997).
\bibitem{Polish} M. Karny {\it et al.}, Nuclear Physics A 640, 3 (1998).
\bibitem{Tain_analysis} J. L. Ta\'in and D. Cano-Ott, Nuclear Instruments and Methods in Physics Research A 571, 728 (2007) and 
Nuclear Instruments and Methods in Physics Research A 571, 719  (2007). 
\bibitem{Cano_response} D. Cano {\it et al.}, Nuclear Instruments and Methods in Physics Research A 430, 333 (1999).
\bibitem{pileup} D. Cano {\it et al.}, Nuclear Instruments and Methods in Physics Research A 430, 488 (1999).
\bibitem{Guadilla_nim} V. Guadilla  {\it et al.}, Nuclear Inst. and Methods in Physics Research, A 910, 79-89  (2018). 
\bibitem{Tain_nsens} J. L. Ta\'in {\it et al.}, Nuclear Instruments and Methods in Physics Research A 774, 17 (2015).
\bibitem{Valencia} E. Valencia {\it et al.}, Phys. Rev. C 95, 024320 (2017).
\bibitem{Guadilla_neutron} V. Guadilla {\it et al.}, Phys. Rev. C 100, 044305 (2019).

\bibitem{Kratz1} K.-L. Kratz {\it et al.}, Z. Phys. A 306, 239 (1982).
\bibitem{Abriola1} D. Abriola and A. A. Sonzogni, Nucl. Data Sheets 107, 2423 (2006).
\bibitem{Tain_nd2010} J. L. Ta\'in {\it et al.}, J. Korean Phys. Soc. 59, 1499 (2011).
\bibitem{Algora-annualreport} A. Algora, ATOMKI Annual Report 2004, p. 17 (2004). 
\bibitem{DTAS} J. L. Ta\'in {\it et al.}, Nuclear Instruments and Methods in Physics Research A 803, 36 (2015).
\bibitem{MTAS} M. Karny {\it et al.}, Nuclear Instruments and Methods in Physics Research A 836, 83 (2016).
\bibitem{Leonid} L. Batist, private communication. 
\bibitem{Jordan-thesis} D. Jordan, PhD thesis, Univ. of Valencia, 2010.
\bibitem{SUN} A. Simon {\it et al.}, Nuclear Instruments and Methods in Physics Research A 703, 16 (2013). 
\bibitem{Rasco} B. C. Rasco {\it et al.}, Phys. Rev. C 95, 054328 (2017).
\bibitem{Rasco_JPS} B. C. Rasco {\it et al.}, JPS Conf. Proc. 6, 030018 (2015).
\bibitem{Spyrou_Oslo} A. Spyrou {\it et al.}, Phys. Rev. Lett. 113, 232502 (2014).
\bibitem{Oslo} M. Guttormsen {\it et al.}, Nucl. Instrum. Methods Phys. Res., Sect. A 255, 518 (1987).
\bibitem{Dombos} A. C. Dombos {\it et al.}, Phys. Rev C 93, 064317 (2016).
\bibitem{Guadilla-thesis} V. Guadilla, Ph.D. thesis, University of Valencia, 2017.
\bibitem{100Tc} V. Guadilla {\it et al.}, Phys. Rev C 96, 014319 (2017).
\bibitem{Huber2} P. Huber and P. Jaffke, Phys. Rev. Lett. 116, 122503 (2016).
\bibitem{235Uenergy} M. F. James, Journal of Nucl. Energy 23, 517 (1969).
\bibitem{WayWigner} K. Way and E. Wigner, Phys. Rev. 73, 1318 (1948).
\bibitem{Rudstam} G. Rudstam {\it et al.}, At. Data Nucl. Data Tables 45, 239 (1989). 
\bibitem{Tengblad} O. Tengblad {\it et al.}, Nucl. Phys. A 503, 136 (1989).
\bibitem{Ensdf} ENSDF, http://www.nndc.bnl.gov/ensdf
\bibitem{WPEC25} T. Yoshida, A. L. Nichols {\it et al.}, Assessment of Fission Product Decay Data for Decay Heat Calculations (OECD/NEA Working Party for International Evaluation Co-operation, Paris, 2007), NEA report NEA/WPEC-25 (2007) 1., Vol. 25 (NEA No. 6284).
\bibitem{Nichols1} A. L. Nichols, IAEA Report No. INDC(NDS) 0499 (2006).
\bibitem{Nichols2} M. Gupta {\it et al.}, IAEA Report No. INDC(NDS) 0577 (2010).
\bibitem{Igisol} J. \"Ayst\"o, Nucl. Phys. A 693, 477 (2001);  I. D. Moore {\it et al.}, Nucl. Instrum. and Methods B 317, 208 (2013).
\bibitem{Ionguide} P. Dendooven, Nuclear Instruments and Methods in Physics Research B 126, 182 (1997).
\bibitem{JYFLTRAP} V. Kolhinen {\it et al.}, Nuclear Instruments and Methods in Physics Research A 528, 776 (2004); T. Eronen {\it et al.}, Eur. Phys. J. A 48, 46 (2012).
\bibitem{Agramunt} J. Agramunt {\it et al.}, Nucl. Data Sheets 120 , 74 (2014); J. Agramunt {\it et al.}, EPJ Web of Conferences 146, 01004 (2017);  
J. Agramunt {\it et al.}, Nuclear Instruments and Methods in Physics Research A 807, 69 (2016).
\bibitem{Jyv1} A. Algora and J. L. Tain, "Beta decay requirements for reactor decay-heat calculations: improving the prediction power of fission products summation calculations", Univ. of Jyv\"askyl\"a IGISOL I116 proposal (2006).
\bibitem{Jyv2} B. Gomez, D. Cano, A. Algora and A. Jokinen, "Decay properties of beta delayed neutron emitters", Univ. of Jyv\"askyl\"a IGISOL I136 proposal (2008).
\bibitem{Jyv3} M. Fallot, J.L Tain and A. Algora, "Study of nuclei relevant for precise predictions of reactor neutrino spectra", Univ. of Jyv\"askyl\"a IGISOL I153 proposal (2009).
\bibitem{Jyv4} A. Algora and J. L. Tain, "Total absorption measurement of the beta-deacy of $^{100}$Tc", Univ. of Jyv\"askyl\"a IGISOL I154 proposal (2009).
\bibitem{ensdf-86Br} A. Negret and B. Singh, Nucl. Data Sheets 124, 1 (2015).
\bibitem{Goriely-ripl3} S. Goriely, F. Tondeur, and J. Pearson, At. Data Nucl. Data Tables 77, 311 (2001); P. Demetriou and S. Goriely, Nucl. Phys. A 695, 95 (2001).
\bibitem{Gamma-stf} R. Capote {\it et al.}, Nucl. Data Sheets 110, 3107 (2009); https://www-nds.iaea.org/RIPL-3/
\bibitem{Rice} S. Rice {\it et al.}, Phys. Rev. C 96, 014320 (2017).
\bibitem{Rice-thesis} S. Rice, PhD thesis, University of Surrey, 2014.
\bibitem{Sonzogni_private} A. Sonzogni, private communication. 
\bibitem{Greenwood_MC} R. C. Greenwood {\it et al.}, Nuclear Instruments and Methods in Physics Research A 314, 514 (1992)
\bibitem{Tain-preprint}  J. L. Ta\'in and A. Algora, IFIC-06-1 Report, 2006.
\bibitem{Zak} A. A. Zakari-Issoufou {\it et al.}, { Phys. Rev. Lett.}  115, 102503 (2015).
\bibitem{Tobias} A. Tobias, CEGB Report No. RD/B/6210/R89, (1989).
\bibitem{Dickens} J. K. Dickens {\it et al.}, Nucl. Science and Eng. 74, 106 (1980) and J. K. Dickens {\it et al.}  Nucl. Science and Eng. 78, 126 (1981).
\bibitem{Giot} L. Giot (Subatech Lab., Nantes, France) private communication. 
\bibitem{Reines} C. L. Jr. Cowan, F. Reines, {\it et al.}, Science 124, 103 (1956).
\bibitem{Mikaelian} L.A. Mikaelian, 1977, Proc. Int. Conf. Neutrino-77, v. 2, p. 383. 
\bibitem{DC} Y. Abe  {\it  et al.}, Phys. Rev. Lett.  108, 131801 (2012).
\bibitem{DayaBay} F. P. An {\it  et al.}, Phys. Rev. Lett.  108, 171803 (2012).
\bibitem{RENO} J. K. Ahn  {\it  et al.}, Phys. Rev. Lett.  108, 191802 (2012). 
\bibitem{Cucoanes2013} A. Cucoanes {\it et al.}, arXiv:1501.00356.
\bibitem{SchreckU5-1} K.~Schreckenbach {\it et al.}, Phys.\ Lett.\ 99B, 251 (1981).
\bibitem{SchreckU5-2} K. Schreckenbach {\it et al.}, Phys.\ Lett.\ 160B, 325 (1985). 
\bibitem{SchreckU5Pu9} F.~von~Feilitzsch, A.~A.~Hahn and K. Schreckenbach, Phys.\ Lett.\ 118B, 162 (1982).
\bibitem{Hahn} A. A. Hahn {\it et al.}, Phys. Lett. B 218, 365 (1989) .
\bibitem{Mueller} T. A. Mueller  {\it et al.}, Phys. Rev.  C 83, 054615 (2011).
\bibitem {Huber} P. Huber, Phys. Rev. C 84, 024617 (2011).
\bibitem{Mention} G. Mention {\it et al.}, Phys. Rev.  D 83, 073006 (2011).
\bibitem{SoLid} L. N. Kalousis and SoLid collaboration, J. Phys.: Conf. Ser. 888, 012181 (2017).
\bibitem{STEREO} H. Almaz\'an {\it et al.} (STEREO Collaboration), Phys. Rev. Lett. 121, 161801 (2018).
\bibitem{Prospect} J. Ashenfelter {\it et al.}, Phys. Rev. Lett. 122, 251801 (2019).
\bibitem{Mention2017} G. Mention {\it et al.}, Phys. Lett.  B 773 (2017) 307, arXiv:1705.09434.
\bibitem{BILL} W.~Mampe {\it et al.},  Nucl.\ Instrum.\ Meth.\ 154, 127 (1978).
\bibitem{OnillonAAP} A. Onillon, "Updated Flux and Spectral Predictions Relevant to the Reactor Antineutrino Anomaly", Applied  Antineutrino Physics Workshop 2018, (2018). 
\bibitem{Hayes} A. C. Hayes, J.L. Friar, G.T. Garvey, G.Jungman, G. Jonkmans, Phys. Rev. Lett. 112, 202501 (2014).
\bibitem{FangBrown} D. L. Fang and B. A. Brown, Phys. Rev. C 91, 025503 (2015).
\bibitem{Hayen} L. Hayen, J. Kostensalo, N. Severijns, and J. Suhonen, Physical Review C 100, 054323 (2019). 
\bibitem{Hayes2017} X.B. Wang, J.L. Friar and A.C. Hayes, Phys. Rev. C 94, 034314 (2016). 
\bibitem{HayesSolvay2017} A. Hayes, contribution to the Solvay workshop (2017).
\bibitem{King} R. W. King and J. F. Perkins, Phys. Rev. 112, 963 (1958).
\bibitem{Avignonne} F. T. Avignone {\it et al.} , Phys. Rev. 170, 931 (1968).
\bibitem{Vogel81} P.~Vogel, G. K.~Schenter, F. M.~Mann and R. E.~Schenter, Phys. Rev. C 24, 1543 (1981).
\bibitem{Haag} N. Haag {\it et al.}, Phys. Rev. Lett.  112, 122501 (2014).
\bibitem{Fallot} M. Fallot {\it et al.}, { Phys. Rev. Lett.} 109, 202504 (2012).
\bibitem{Sonzo2015} A. A. Sonzogni, T. D. Johnson, and E. A. McCutchan, Phys. Rev. C 91, 011301(R) (2015).
\bibitem{Rasco_let} B. C. Rasco {\it et al.}, Phys. Rev. Lett. 117, 092501 (2016); 
\bibitem{Alexa} A. Fija\l{}kowska {\it et al.}, Phys. Rev. Lett. 119, 052503 (2017).
\bibitem{Report-NDS-676} P. Dimitriou and A. L. Nichols, { IAEA report INDC(NDS)-0676}, Feb. 2015, IAEA, Vienna, Austria
\bibitem{Dwyer} D. A. Dwyer and T. J. Langford, Phys. Rev. Lett. 114, 012502 (2015).
\bibitem{Jeff} JEFF and EFF projects http://www.oecdnea.org/dbdata/jeff/, URL http://www.oecd-nea.org/dbdata/jeff/.
\bibitem{endf} D. Brown {\it et al.}, Nucl. Data Sheets 148, 1 (2018).  
\bibitem{Magali_priv} M. Estienne (Subatech lab., Nantes, France) private communication.
\bibitem{DayaBay2017} F. P. An {\it et al.} (Daya Bay Collaboration), Phys. Rev. Lett. 118, 251801 (2017).
\bibitem{Hayes2018} A. C. Hayes {\it et al.}, Phys. Rev. Lett. 120, 022503 (2018).
\bibitem{Estienne2019} M. Estienne  {\it et al.}, Phys. Rev. Lett. 123, 022502 (2019).
\bibitem{GrossTheo2018} T. Yoshida, T. Tachibana, S. Okumura, and S. Chiba, Phys. Rev. C 98, 041303(R) (2018).
\bibitem{BeacomVogel} P. Vogel and J. F. Beacom, Phys. Rev.  D 60 (4), 053003 (1999).
\bibitem{DayaBay2016} F. P. An  {\it et al.}, Chinese Phys. C 41, 013002 (2017).
\bibitem{NEOS} Y. J. Ko {\it et al.},  Phys. Rev. Lett. 118, 121802 (2017). 
\bibitem{Bugey} B. Achbar {\it et al.}, Phys. Lett. B 374 243  (1996). 
\bibitem{DCIV} H. de Kerret {\it et al.},  Nat. Phys. 16, 558 (2020).
\bibitem{Berryman} J. M. Berryman and P. Huber, Phys. Rev. D 101, 015008 (2020).
\bibitem{Sonzo2018} A. A. Sonzogni {\it et al.},  Phys. Rev. C 98, 014323 (2018).
\bibitem{Juno} E. Ciuffoli, J. Evslin and X. Zhang, Phys. Rev. D 88 (3) 033017 (2013).
\bibitem{Juno-Tao}  An F {\it et al.},  J. Phys. G: Nucl. Part. Phys. 43 030401 (2016). 
\bibitem{COHERENT} D. Akimov {\it et al.}, Science  Vol. 357, Issue 6356, 1123 (2017).
\bibitem{Rubio_euroschool} B. Rubio and W. Gelletly, {\it Beta Decay of exotic nuclei}  The Euroschool Lectures on Physics with Exotic Beam Vol III (Springer Lecture Notes in Physics vol 764, year 2009, p- 99) ed. J.S. Al-Khalili and E. Roeckl (Berlin Springer).
\bibitem{Krane} K. S. Krane, {\it Introductory Nuclear Physics} (Wiley, New York, 1988).
\bibitem{Evans} R. D. Evans, {\it The Atomic Nucleus} (McGraw-­‐Hill, New York, 1956).
\bibitem{Siegbahn} K. Siegbahn, {\it Alpha, Beta and Gamma Ray Spectroscopy} (North-­‐Holland, Amsterdam, 1966).
\bibitem{Ichimura}M. Ichimura, H. Sakai and T. Wakasal Prog. Part. Nucl. Phys 56, 451 (2006).
\bibitem{Pinedo} G. Martinez-Pinedo, A. Poves, E. Caurier and A. P. Zuker, Phys. Rev. C 53, R2602 (1996).  
\bibitem{Nature_quenching} P. Gysbers {\it et al.}, Nature Physics 15, 428 (2019).
\bibitem{Towner}  I. S. Towner. Nucl. Phys. A 444, 402 (1985).
\bibitem{MSEP} K.H. Burkard {\it et al.}, Nucl. Instrum. Methods 139, 275 (1978).
\bibitem{KarnyTAS} M. Karny {\it et al.}, Nucl. Instr. Methods B 126, 411 (1997).
\bibitem{Plettner} C. Plettner {\it et al.}, Phys. Rev. C 66, 044319 (2002).
\bibitem{Hu} Z. Hu {\it et al.}, Phys. Rev. C 60, 024315 (1999).
\bibitem{Algora_Dy} A. Algora {\it et al.}, Phys. Rev. C 70, 064301 (2004).
\bibitem{Algora_clus} A. Algora {\it et al.}, Phys. Rev. C 68, 034301 (2003).
\bibitem{Cano_150} D. Cano-Ott, Ph.D. thesis, University of Valencia, 2000.
\bibitem{Nacher2} E. Nacher {\it et al.}, Phys. Rev. C 93 014308 (2016).
\bibitem{Hinke} C. B. Hinke {\it et al.}, Nature 496, 341 (2012).
\bibitem{Lubos} D. Lubos {\it et al.}, Phys. Rev. Lett. 122, 222502 (2019).
\bibitem{RIKEN_Proposal} A. Algora and B. Rubio, " Studies of the beta decay of $^{100}$Sn and its neighbours with a Total Absorption Spectrometer (TAS)", RIKEN proposal NP1612-RIBF147 (2016).
\bibitem{Algora-annualreport2} A. Algora {\it et al.}, RIKEN Progress Report 2019, in print.
\bibitem{Batist} L. Batist {\it et al.}, Eur. Phys. J. A 46, 45 (2010).
\bibitem{Hamamoto} I. Hamamoto and X. Z. Zhang, Z. Phys. A 353, 145 (1995).
\bibitem{Sarriguren} P. Sarriguren  {\it et al.}, Nucl. Phys. A 635, 55 (1998); P. Sarriguren  {\it et al.},  Nucl. Phys. A 658 13 (1999); P. Sarriguren  {\it et al.}, Nucl. Phys. A 691, 631 (2001).
\bibitem{Petrovici} A. Petrovici, A. Schmid and A. Faessler, Nucl. Phys. A 665 333 (2000).
\bibitem{Guadilla_PRCNb} V. Guadilla {\it et al.}, Phys. Rev. C 100, 024311 (2019).
\bibitem{Guadilla_inprep} V. Guadilla {\it et al.}, in preparation.
\bibitem{Sarriguren_priv} P. Sarriguren, private communication.
\bibitem{Kratz-Moller} K. L. Kratz and P. M\"oller private communication.
\bibitem{Jordan-2} D. Jordan {\it et al.} in preparation.
\bibitem{Watson} D. Watson {\it et al.}, Nature 574, 497 (2019).
\bibitem{Horowitz} C. J. Horowitz {\it et al.}, J. Phys. G: Nucl. Part. Phys. 46, 083001 (2019).
\bibitem{Tain_Mazurian} J.L. Tain {\it et al.}, Acta Phys. Polonica B 49, 417 (2018).
\bibitem{Caballero} R. Caballero-Folch {\it et al.}, Phys. Rev. Lett. 117, 012501 (2016).
\bibitem{Larsen} A.C. Larsen {\it et al.}, Prog. Part. Nucl. Phys. 107, 69 (2019).
\bibitem{Hauser} W. Hauser and H Feshbach, Phys. Rev. 87, 366 (1952).
\bibitem{Spyrou} A. Spyrou {\it et al.}, Phys. Rev. Lett. 117, 142701 (2016).
\bibitem{Gottardo} A. Gottardo {\it et al.}, Physics Letters B772, 359  (2017).
\bibitem{Mumpower} M. R. Mumpower {\it et al.}, Phys. Rev. C 94, 064317 (2016).
\bibitem{Horowitz2} C. J. Horowitz {\it et al.}, Phys. Rev. Lett. 86, 5647 (2001). 
\bibitem{Piekarewicz} J. Piekarewicz, Phys. Rev. C 73 044325 (2006). 
\bibitem{Goriely} S. Goriely, Phys. Lett. B 436, 10 (1998). 
\bibitem{Alto_prop} M. Fallot, A.Porta, A. Algora, B. Rubio and J.-L. Tain, "Total Absorption Spectroscopy for Nuclear Structure, Nuclear Astrophysics, Neutrino and Reactor Physics", ALTO facility Proposal Ref. N-RI-6 (2014).  
\bibitem{Scheck} M. Scheck {\it et al.}, Phys. Rev. Lett. 116, 132501 (2016).
\end{thebibliography}
%

\end{document}